\documentclass[a4paper,11pt]{article}

\usepackage{jcappub} 
\usepackage{subfigure}
\usepackage{bm}

\usepackage[T1]{fontenc} 

\newcommand{\rmnum}[1]{\uppercase\expandafter{\romannumeral #1\relax}}
\newcommand{\bea}{\begin{eqnarray}}
	\newcommand{\eea}{\end{eqnarray}}
\newcommand{\beq}{\begin{equation}}
	\newcommand{\eeq}{\end{equation}}
\newcommand{\nn}{\nonumber}
\newcommand{\llangle}{\left\langle}
\newcommand{\rrangle}{\right\rangle}

\begin{document}

\title{\boldmath The effects on CMB power spectra and bispectra from the polarization rotation and its correlations with temperature and E-polarization} 	

\author[a,b,1]{Hua Zhai,\note{Corresponding author.}}
\author[c]{Si-Yu Li,}
\author[d,e]{Mingzhe Li,}
\author[c]{Hong Li,}
\author[a,b]{and Xinmin Zhang.}

\affiliation[a]{Theoretical Physics Division, Institute of High Energy Physics (IHEP), Chinese Academy of Sciences, 19B Yuquan Road, Shijingshan District, Beijing 100049, China}
\affiliation[b]{University of Chinese Academy of Sciences, Beijing, China}
\affiliation[c]{Key Laboratory of Particle Astrophysics,  Institute of High Energy Physics (IHEP), Chinese Academy of Sciences, 19B Yuquan Road, Shijingshan District, Beijing 100049, China}
\affiliation[d]{Interdisciplinary Center for Theoretical Study, University of Science and Technology of China, Hefei, Anhui 230026, China}
\affiliation[e]{Peng Huanwu Center for Fundamental Theory,  Hefei, Anhui 230026, China}

\emailAdd{zhaihua@ihep.ac.cn}
\emailAdd{lisy@ihep.ac.cn}
\emailAdd{limz@ustc.edu.cn}
\emailAdd{hongli@ihep.ac.cn}
\emailAdd{xmzhang@ihep.ac.cn}

\abstract{The Chern-Simons term, through which the cosmic Axion-like field couples to the electromagnetic field, has the effect to rotate CMB polarization directions and to break the CPT symmetry. This rotation will change the CMB power spectra, no matter isotropic or anisotropic the rotation angle is. In this paper we revisit this issue by further considering the correlations between the anisotropic rotation angle $\alpha$ and the CMB temperature and CMB $E$ polarization fields. These correlations could be generated in the Axion-like models with nonzero potential under the adiabatic initial condition. We first investigate how these correlations contribute further modifications to the CMB power spectra, then calculate the CMB bispectra for the temperature and rotated polarization fields. These bispectra would vanish if the $T\alpha$ and $E\alpha$ correlations are absent. So, they are useful in searching for CPT violation and the $T\alpha$ and $E\alpha$ correlations arisen in the Axion-like models.}

\maketitle
\flushbottom

\section{Introduction}
The Charge-Parity-Time Reversal (CPT) symmetry is exact and occupies a fundamental status in the standard model of particle physics. Since decades people realized that searching for possible CPT violations is an important way to test the standard model and a convenient approach to  seek the new physics. Up to now, the CPT symmetry has passed a number of high-precision experimental tests and no definite signal of its violation has been observed in the laboratory. So, the  CPT violation, if exist, should be very small to be amenable to the laboratory experimental limits.

However, the CPT symmetry could be dynamically broken in the expanding universe.  For instances, in refs. \cite{Cohen:1987vi, Li:2001st,Li:2002wd,Davoudiasl:2004gf,Li:2004hh}, the cosmological CPT violation has been considered to generate the baryon number asymmetry in the early universe. A notable property of this kind of baryogenesis models is that the CPT violation at present time is too small to be detected by the laboratory experiments, but was large enough in the early universe to account for the  observed baryon number asymmetry.  As shown in refs. \cite{Feng:2004mq,Feng:2006dp,Li:2006ss}, such type of CPT violations might be observed by the cosmological probes. With the accumulation of high-quality observational data, especially those from the cosmic microwave background (CMB) experiments, cosmological observation becomes a powerful way to test CPT symmetry.  

Generally the cosmological CPT violation in the photon sector can be modeled by the coupling between photons and an external field $\theta(x)$ through the Chern-Simons Lagrangian,
\bea\label{cs}
\mathcal{L}_{cs} = \theta(x) F_{\mu\nu} \widetilde{F}^{\mu\nu}~,
\eea
where $F_{\mu\nu}=\partial_{\mu}A_{\nu}-\partial_{\nu}A_{\mu}$ is the electromagnetic tensor and  $\widetilde{F}^{\mu\nu}=(1/2)\epsilon^{\mu\nu\rho\sigma}F_{\rho\sigma}$ is its dual. If  $\theta$ is a constant, the Chern-Simons term will have no effect on the dynamics of photons because the Pontryagin density $F_{\mu\nu} \widetilde{F}^{\mu\nu}$ is a total derivative and the vacuum here is topological trivial. There are at least two approaches to get $\theta(x)$ as a variable. With the first approach, $\theta(x)=p_{\mu}x^{\mu}$ is constructed by a non-dynamical vector $p_{\mu}$. However when considering the couplings to gravity, this case is not compatible with general relativity and its covariant extensions \cite{Li:2009rt}. In the second approach, $\theta(x)=f(\phi(x))$,where $f(\phi(x))$ is a general function of a dynamical scalar field $\phi(x)$.  Such a scalar field may be the dynamical dark energy \cite{Wetterich:1987fm,Caldwell:1999ew,Feng:2004ad} as in refs. \cite{Li:2001st,Li:2002wd}, or Axion-like field, or the curvature of spacetime \cite{Davoudiasl:2004gf,Li:2004hh}. During the evolution of the universe,  $\theta(x)$ is treated as an external field, its evolution or configuration picks up a preferred frame, so that the Chern-Simons term (\ref{cs}) is not invariant under Lorentz and CPT transformations. The physical consequence of this CPT violation is to cause the rotations of the polarization directions of photons when propagating in the space. This holds for both the homogeneous and  inhomogeneous universe \cite{Li:2008tma}. In terms of the Stokes parameters for linear polarized photons, the rotation can be expressed as 
\beq\label{rotate_stoke} 
\widetilde Q \pm i \widetilde  U = \exp(\pm	i 2\alpha)(Q\pm i U)~,
\eeq
where the rotation angle is twice the integral of $\partial_{\mu}\theta$ along the light ray from the source to the observer,
\bea\label{alpha}
\alpha =2 \int_s^o  \partial_{\mu}\theta dx^{\mu}=2[\theta(x_o)-\theta(x_s)]\equiv 2\Delta \theta~,
\eea
and finally proportional to the change of $\theta$ over the photon trajectory. 
For CMB, the photons we received today came from last scattering surface (LSS) at which they decoupled with matter. It is convenient for us to set the observer at the origin point of the coordinate system so that 
\bea
x_0=(\eta_0, \bm 0)~,~x_s=( \eta_{lss}, -\Delta\eta {\bm n}),
\eea 
where $\bm{n}$ represents the propagating direction of CMB photon from the point at LSS and finally received by the observer, $\Delta \eta=\eta_0-\eta_{lss}$ is the conformal time difference. 
This rotation has the ability to convert part of CMB E-mode polarization to B-mode polarization, and vice versa. This will change the power spectra of CMB polarization, especially induce nonzero TB and EB spectra \cite{Lue:1998mq,Feng:2006dp}. Such effects offer a way to detect or constrain the rotation angle, then the CPT-violation signature with CMB data.

The rotation angle $\alpha(\bm{n})$ is generally a direction dependent scalar field on 2-d sphere, as shown in eq.~(\ref{alpha}). It is natural to split  $\alpha(\bm{n})$ into the isotropic and the anisotropic parts $
\alpha(\bm{n})=\bar{\alpha} + \delta\alpha(\bm{n})$ as we have usually done in the cosmological perturbation theory.  The isotropic rotation angle, $\bar{\alpha}$, can be considered as the mean of $\alpha(\bm{n})$ over the sphere.  At the leading order, we may only consider the isotropic rotation angle as an approximation.  For this case, the rotated CMB power spectra have simple forms \cite{Feng:2006dp}. With WMAP and BOOMERANG (B03) data, Feng et.al \cite{Feng:2006dp} has performed the first measurement on the isotropic rotation angle. Since then, a lot of works have been done in terms of the observed CMB polarization data along this line. It has been constrained by various collaborations of CMB surveys, including QUaD \cite{Wu:2008qb}, WMAP \cite {Hinshaw:2012aka}, ACTPol \cite{Louis:2016ahn} and Planck \cite{Aghanim:2016fhp}, and by combined datasets including CMB and LSS observations \cite{Xia:2009ah, Zhao:2015mqa, Xia:2012ck}. The constraint on $\bar{\alpha}$ in these works is found to be at the level of one degree. Up to now the Planck collaboration gave the most stringent limit  \cite{Aghanim:2016fhp} $\bar{\alpha} = 0.35^\circ \pm 0.33^\circ$. 

A comprehensive study on the Chern-Simons effect should include the spatial dependence or the anisotropies of the rotation angle \cite{Li:2008tma}. If the anisotropies are random and satisfy the Gaussian statistics, they can be described fully by an angular power spectrum $C_{l}^{\alpha\alpha}$ .  In terms of $C_{l}^{\alpha\alpha}$,  refs. \cite{Li:2008tma, Li:2013vga} derived the analytic formulae of distortion effects on CMB power spectra, and then refs. \cite{Li:2013vga, Li:2014oia} constrained the anisotropies by global fitting to the combination of CMB observations.
Anisotropic rotation angles are also studied in refs. \cite{Kamionkowski:2008fp, Yadav:2009eb, Gluscevic:2009mm}  using the four point correlation function  method.  In this way, the constraints on the anisotropies of rotation angle with WMAP-7 \cite{Gluscevic:2012me}, POLARBEAR \cite{Ade:2015cao}, and BICEP2/Keck Array \cite{Array:2017rlf} were obtained.
To date the most stringent limit is $l(l+1)/2\pi C_l^{\alpha\alpha} \le  0.033 ~\mathrm{deg}^2$ by the ACTPol experiment \cite{Namikawa:2020ffr}.  Unlike the isotropic rotation angle,  $\delta\alpha(\bm n)$ does not predict significant signal on $TB$ and  $EB$ power spectra \cite{Zhai:2019god} but both of therm will produce ambiguous $BB$  power spectrum and affect the detection of primordial gravitational waves.  Estimating results of $C_l^{\alpha\alpha}$ from the $BB$ power spectrum \cite{Li:2013vga, Li:2015vea, Zhai:2019god} are consistent with the results obtained from reconstruction method \cite{Gluscevic:2012me, Ade:2015cao,Array:2017rlf}.  

In previous studies, the anisotropic rotation angle $\delta\alpha(\bm n)$ was assumed to be uncorrelated with the primordial temperature and polarization fields at LSS. The cross-correlation of the rotation angle with temperature was first considered in ref. \cite{Caldwell:2011pu} with the model where the $\theta(x)$ field which coupled to the Pontryagin density in the Chern-Simons Lagrangian (\ref{cs}) is originated from the cosmic Axion-like field with non-vanishing potential,
\bea\label{quintessencecs}
\mathcal{L} _{cs}= \frac{\beta \phi}{2M} F^{\mu\nu} \widetilde{F}_{\mu\nu}~,~{\rm and}~
\alpha = \frac{\beta}{M}\Delta\phi~,
\eea  
where $M$ is a mass scale and $\beta$ is the dimensionless coupling constant. Recently similar $T\alpha$ correlation was also considered in ref. \cite{Capparelli:2019rtn} with an early dark energy model which was proposed to resolve the Hubble tension \cite{Poulin:2018cxd}.  In this paper, we will also take the Axion-like field as an example and will take both $T\alpha$ and $E\alpha$ cross-correlations into account. It is not necessary to consider $B\alpha$ correlation, because $\delta\alpha(\bm n)$ is a scalar perturbation induced by $\delta \phi$, in linear perturbation theory it is expected to be uncorrelated with the primordial $B$ mode polarization which was seeded by the primordial gravitational waves. We will first investigate how the $T\alpha$ and $E\alpha$ cross-correlations change the power spectra of CMB and then focus on the bispectra of the rotated CMB polarization field. 

Nonzero bispectrum or three point correlation function means non-Gaussian statistics. In this paper, we assume the temperature and the unrotated polarization fields of CMB and the anisotropic rotation angle are all Gaussian random fields. 
So the phase factor $\exp(\pm 2i \alpha)$ in eq.~(\ref{rotate_stoke})  has a  log-normal distribution. This may cause deviations from Gaussian distribution for the rotated CMB polarizations.  Due to $T\alpha$ and $E\alpha$ correlations, the rotated three point function of CMB, $\llangle \widetilde{a}_{l_1 m_1}^{X_1} \widetilde{a}_{l_2 m_2}^{X_2} \widetilde{a}_{l_3 m_3}^{X_3} \rrangle $, is actually the four point correlation function by $X_i$ and exponent function of $\alpha$, here at least one $X_i$ is the polarization field. If the unrotated CMB field have nonzero correlations with $\delta\alpha$ , then the three point function will be nonzero even if there is no other non-Gaussianity in the unrotated three point correlation function, 
$\llangle {a}_{l_1 m_1}^{X_1} {a}_{l_2 m_2}^{X_2} {a}_{l_3 m_3}^{X_3} \rrangle $. Thus, the bispectra  for the rotated CMB polarization field are not only important to search for the CPT violation, but also essential for the $T\alpha$ and $E\alpha$ cross-correlations.

However the unrotated three point function may be nonzero when there are other non-Gaussian sources. Primordial parity-even bispectra $ TTT,   TTE,   TEE,   EEE$ with $l_1+l_2+l_3= $ even configuration are predicted in various inflation models \cite{Maldacena:2002vr,Komatsu:2001rj,  Bartolo:2004if, Baumann:2009ds} or models alternative to inflation \cite{Cai:2009fn, Cai:2010kp}. Parity violation can also arise during inflation in the frame of primordial gravitational wave \cite{Maldacena:2011nz, Soda:2011am, Shiraishi:2011st}, whereas parity-odd bispectra with $l_1+l_2+l_3= $ odd configuration are generated for  $T$ and $E$.  The primordial bispectra are usually characterized by a non-linear parameter $f^{\mathrm{loc}}_{NL}$ \cite{Komatsu:2001rj} for even parity or $f_{NL}^{\mathrm{ten}}$ \cite{Shiraishi:2014roa, Shiraishi:2014ila,  Akrami:2019izv} for odd parity. In addition, the non-Gaussianity generated by primordial magnetic field was also studied \cite{Shiraishi:2013vha}. Except for the primordial origins,  non-Gaussianity may be  generated by  the late time secondary effects such as the correlations between the Integrated Sachs-Wolfe effect and weak lensing potential \cite{Lewis:2011fk}.  In addition, the higher order of cosmological perturbation theory can also give rise to non-Gaussinities \cite{Pitrou:2008hy}. Given these studies,  non-Gaussianity of CMB can be powerful probe for the early universe theories and the late time evolution. To date,  Planck \cite{Akrami:2019izv} analyzes the non-Gaussianity from  CMB temperature and E polarization map and gives stringent limits $f_{NL}^{\mathrm{loc}} = -0.9 \pm 5.1$ for parity-even bispectra, and $f_{NL}^{\mathrm{ten}}  = (1 \pm 18) \times 10^2$  for parity-odd ones. The results show no significant non-Gaussianity signature and thus put strong constraints on various theories. In this paper, for simplicity we will not consider the mixture of the polarization rotation effect with other non-Gaussianity sources, but take them as comparisons. 

The structure of this paper is organized as follows. In section \ref{quintessence_model}, we review how $T\alpha$ and $E\alpha$ cross power spectra are produced by the Chern-Simons term in Axion-like model. In section \ref{section_power_spectrum}, we obtain the rotated CMB power spectra results when $C_l^{T\alpha}$ and $C_l^{E\alpha}$ are considered and compare them with previous results where these cross correlations were ignored. In section \ref{section_bispectrum}, we derive and analyze the bispectra results of the rotated CMB fields. We point out that the bispectra can only be produced with non-zero $C_l^{T\alpha}$ and $C_l^{E\alpha}$. Section \ref{conclusion} is dedicated to the conclusion and discussions. Some detailed calculations and mathematical tools can be found in the appendices.  

\section{Anisotropic rotation angles in context of Axion-like model}
\label{quintessence_model}

In this section, we make a short review of the statistical description of anisotropic polarization rotation angle and then give the theoretical formulae of cross correlation between it and CMB temperature and polarization in frame of an Axion-like scalar field. Although other scalar models may also allow nonzero cross correlations, Axion scalar model are more widely used in current dark energy and dark matters issues.

In the model of Axion-like scalar field that is coupled to the electromagnetic field through the Chern-Simons term (\ref{quintessencecs}), the rotation angle for CMB polarization direction is 
\bea
\alpha=\frac{\beta}{M}[\phi(x_0)-\phi(x_{lss})]~.
\eea
this angle is induced by a dynamical field and is generally anisotropic because the Axion-like field $\phi$ is not homogeneously distributed in the universe. As usually done in the cosmological perturbation theory, 
the rotation angle can be split into the isotropic part and a zero mean anisotropic part,
\bea
\alpha(\bm n) = \bar{\alpha}+\delta\alpha(\bm n), ~~ \llangle \delta\alpha(\bm n) \rrangle =0~.
\eea
the isotropic rotation angle corresponds to $\bar{\alpha}=(\beta/M)[\bar{\phi}(\eta_0)-\bar{\phi}(\eta_{lss})]$, determined by the background evolution of the Axion-like field. 
The anisotropic rotation angle reads 
\bea
\delta\alpha(\bm n) = \frac{\beta}{M}\left[   \delta\phi(\eta_0, \bm x_0) - \delta\phi(\eta_{lss}, \bm x_{lss})  \right]~.
\eea 
The first term at the right hand side  only contributes an unobservable monopole which acts like the monopole part of CMB fluctuation~\cite{Dodelson:2003ft}. Because $ \delta\phi(\eta_0, \bm x_0)$ depends on the observer's position, it cannot be absorbed into the definition of the isotropic
rotation angle. In this paper we will neglect it as well as the dipole term.  
Using the relation $\bm x_{lss}=-\Delta\eta \bm n$, we have 
\bea\label{relation1}
\delta\alpha(\bm n) =- \frac{\beta}{M}\delta\phi(\eta_{lss}, -\Delta\eta \bm n) ~,
\eea 
the anisotropy of the rotation angle depends on the distribution of the Axion-like scalar field on the last scattering surface. 

As a scalar on the 2-d sphere, the anisotropic rotation angle is usually expanded with the spherical harmonics,
\bea \label{alphaex}
\delta\alpha(\bm n) =\sum_{l m}\alpha_{l m} Y_{lm}(\bm n)~.
\eea
The relation (\ref{relation1}) implies that  \cite{Li:2008tma} ,
\bea
\alpha_{l m}&=&  -  \frac{1}{2\pi^2}(-i)^{l}\frac{\beta}{M}\int d^3\bm k \delta\phi(\bm k, \eta_{lss})j_{l}(k\Delta\eta)Y^{\ast}_{lm}( \hat{\bm k} ),
\label{anisotropy_harmonic_general}
\eea
where $\delta\phi(\bm k, \eta_{lss})$ is the Fourier transformation of  $\delta\phi(\eta_{lss}, -\Delta\eta \bm n) $, $j_{l}$ is the spherical Bessel function and $\hat{\bm k}$ is unit of the vector $\bm k$ in Fourier space.

The perturbation equation for the Axion-like field is 
\bea\label{perturbationeq}
\delta\ddot{\phi}+ 2 \mathcal{H} \delta\dot{\phi}+a^2V''\delta\phi+k^2\delta\phi=\dot\phi(3\dot\Phi+\dot\Psi)-2a^2V'\Psi~, 
\eea	 
in the conformal-Newtonian gauge, where dot means derivative with respect to the conformal time, prime means derivative of the potential to the scalar field, $\mathcal{H}=\dot a/a$ is the conformal Hubble parameter, $\Phi$ and $\Psi$ are metric perturbations and they equal to each other if we neglect the anisotropic stress induced by radiation. At the matter dominant epoch, $\Phi$ and $\Psi$ are constants, so we can see if the scalar field is massless, $V=0$, its perturbation decouples from the metric perturbation and its dynamical equation (\ref{perturbationeq}) is homogeneous, the solutions to it remain constant or decaying at super-horizon scales. As Axion-like model considered in this paper, the scalar field should have a nonzero potential, with the couplings like the Chern-Simons term. And the potential is given as follows, 
\bea
V(\phi) = \Lambda^4(1-\cos\frac{\phi}{f})~,
\eea
this potential gives $\phi$ an effective mass $m\sim \Lambda^2/f$. Usually the scale $f$ is much higher than $\Lambda$ and this yields a small mass for the scalar field. 

Given the Axion-like potential, eq.~(\ref{perturbationeq}) is inhomogeneous, like that of a forced oscillator, its solution is the combination of the solution to the homogeneous equation and a special solution for the full inhomogeneous one. The former corresponds to the mode of entropy (or isocurvature) perturbation and the latter is the adiabatic one. What we are interested in here is the adiabatic perturbation, which is seeded by the metric perturbation and the main contribution to it comes from dark matter. Because the same metric perturbation seeded the temperature anisotropy and $E$ mode polarization of CMB at LSS, the perturbation $\delta\phi$ should correlate with $T$ and $E$, it must generate the $T\alpha$ and $E\alpha$ correlations. 

Under the adiabatic initial condition, a solution for $\delta \phi$ was obtained in synchronous gauge under the slow-rolling approximation \cite{Caldwell:2011pu}
\bea
(\delta\phi)_{\mathrm{syn}}(\bm k, \eta_{ lss}) = -\frac{2}{9}\left( \frac{3\Omega_\phi(\eta_{lss}) (1+w_\phi(\eta_{lss}))}{8\pi}\right)^{1/2}M_{PL}\Psi(\bm k, \eta_{lss})~,
\label{quintessence_solution} 
\eea
where $\Omega_{\phi}$ and $w_{\phi}$ are the density parameter and equation of state of $\phi$ at LSS, and $M_{PL}=1/\sqrt{G}$ is the Planck mass. 

In terms of eq.~(\ref{anisotropy_harmonic_general}), and by defining a parameter
\bea
\epsilon = \frac{1}{100} \cdot \frac{1}{9\pi}\frac{\beta M_{PL}}{M}\left( \frac{3\Omega_\phi(\eta_{lss}) (1+w_\phi(\eta_{lss}))}{8\pi}\right)^{1/2}\bigg/\left[\mathrm{rad}\right]~,
\eea
we have the angular power spectrum for the anisotropic rotation angle $C_l^{\alpha\alpha}$,
\begin{align}
C_l^{\alpha\alpha}
&=8\times 10^4 \pi  \epsilon^2 \int k^2 {d k} P_\Psi(k) \left[ j_{l}(k\Delta\eta) T(k, \eta_{lss})\right]^2~~\left[\mathrm{rad}^2\right],
\end{align}
where $T(k, \eta_{lss})$ is the transfer function which evolves the perturbation from primordial to LSS. The definition of the angular power spectrum can be found in the next section. 

The parameter $\epsilon$ has the meaning of square root of the variance of the anisotropic rotation angles \cite{Caldwell:2011pu}. At recombination epoch, the transfer function $T(k, \eta_{lss})  \simeq  1$ at large scale limit $l \le 100$. For scale invariant primordial power spectrum  
$P_\Psi(k) =  \frac{9}{25}\frac{2\pi^2}{k^3} A_s$, the power spectrum of $\delta\alpha$ has the approximate form 
\bea
C_l^{\alpha\alpha} &=&  \frac{7.2\times 10^5\pi^3 \epsilon ^2 A_s }{25 l (l+1)}~~\left[\mathrm{rad}^2\right].
\label{cl_aa}
\eea
when taking $A_s= 2.10\times 10^{-9}$ \cite{Akrami:2018odb}, we approximately have $D_l^{\alpha\alpha} =l(l+1)C_l^{\alpha\alpha}/2\pi \approx 3.0\times 10^{-4} \epsilon^2~ \mathrm{rad}^2$. Current result from ACTPol \cite{Namikawa:2020ffr}:  $D_l^{\alpha\alpha} < 1.0\times 10^{-5} \mathrm{rad}^2$ at 95\% CL, puts a constraint on parameter: $ \epsilon < 0.18$ at 95\% CL. 

Since $\delta\phi$ was sourced by the metric perturbations which also seeded the CMB temperature and $E$ mode polarization anisotropies at LSS, there should exist $T\alpha$ and $E\alpha$ cross-correlations. Recall that \cite{Ma:1995ey}   
\bea
a^T_{lm} &=& \frac{1}{{2\pi^2}}(-i)^{l}\int d^3 \bm{k}  \Psi(\bm k)  \Delta_{Tl}( k, \eta)  Y^*_{lm}(\hat{\bm k})~, \nn\\
a^E_{lm} 
&=&  \frac{1}{{2\pi^2}} (-i)^{l}\int d^3\bm{k}  \Psi(\bm k)  \Delta_{El}( k, \eta)  Y^*_{lm}(\hat{\bm k})~,
\eea
where $\Delta_{Tl}( k, \eta), \Delta_{El}( k, \eta)$ are the transfer functions for temperature and $E$ polarization respectively. We then have the cross-correlation spectra,
\begin{align}
C_l^{T\alpha}&=  4\times 10^2 \epsilon \int k^2d k ~P_\Psi(k) \Delta_{Tl}( k, \eta)j_{l}(k\Delta\eta) T(k, \eta_{lss})~~\left[\mu \mathrm{K}\cdot \mathrm{rad}\right],   \\
C_l^{E\alpha }&=  4\times 10^2 \epsilon \int k^2d k ~P_\Psi(k) \Delta_{El}( k, \eta)j_{l}(k(\Delta\eta) T(k, \eta_{lss})~~\left[\mu \mathrm{K}\cdot \mathrm{rad}\right],
\label{cl_ea}
\end{align}

\begin{figure}[t]
	\centering
	\includegraphics[trim=5cm 0 0 0 0, width=1.08\textwidth]{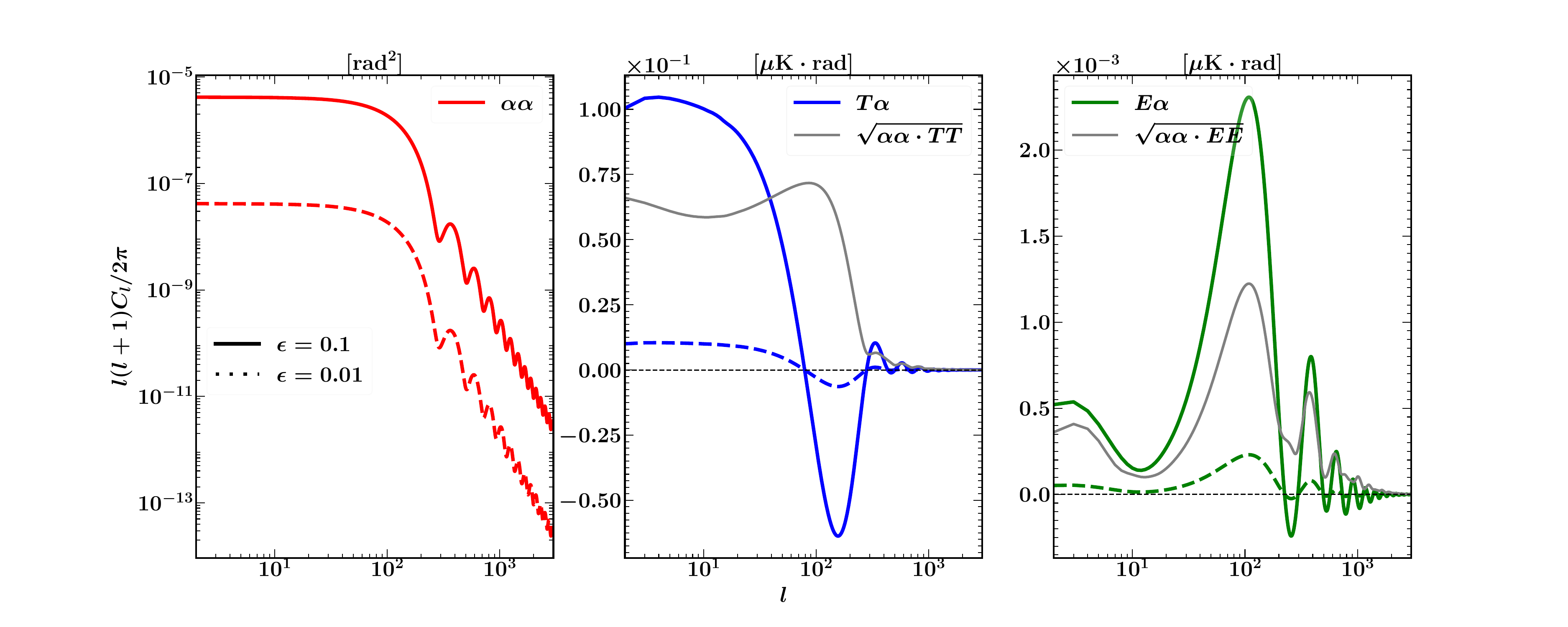}
	\caption{The auto power spectra of anisotropic rotation angle and its correlations with the CMB temperature and $E$-polarization field in Axion-like model, calculated by modifying the CAMB package~\cite{Lewis:1999bs}.  In each panel, the solid and dashed  curves correspond to parameter $\epsilon= 0.1, 0.01$.  The $TT$ and $EE$ power spectra are calculated with Planck 2018 best fit $\Lambda$CDM model. In right two panels,  $C_l^{T\alpha}$ pass zero at $l \approx 80$, while  $C_l^{E\alpha}$ has a cross point at $l \approx 200$. For comparisons,  $l(l+1)\sqrt{C_l^{\alpha\alpha} C_l^{TT}}/2\pi$ and $l(l+1)\sqrt{C_l^{\alpha\alpha} C_l^{EE}}/2\pi$ are plotted with $\epsilon=0.1$.}
	\label{aa_power}	
\end{figure}

Currently there is no direct constraint on these cross-correlations. We investigate in the rest of this paper how the cross-correlations change the power spectra and generate the bispectra of the rotated CMB polarizations, in order to find a way to probe these correlations. Before that, we show the properties of these cross correlations vividly. In figure~\ref{aa_power}, all those power spectra of anisotropic rotation angle similarly have significant values at small multipoles and decrease rapidly with $l$.  The auto power spectra $C_l^{\alpha\alpha}$ has a scale invariant feature at $l \le 100$. It oscillates fast when $l$ increases, so as for $C_l^{T\alpha}$,  $C_l^{T\alpha}$. For numerical sake,  we make a cut  at multipole $l \approx 1000$ for $C_l^{\alpha\alpha}$, and  $l \approx 2000$ for the cross-correlations in the computations. 

\section{Power spectra of rotated CMB polarization fields modified by $C_l^{T\alpha}$ and $C_l^{E\alpha}$}
\label{section_power_spectrum}

The effect of polarization rotation angle on CMB polarization power spectra has already been studied by refs. (\cite{Li:2008tma, Li:2013vga}), but in these previous works the cross correlation between $\delta\alpha$ and CMB temperature as well as E-mode polarizations were ignored. In this section, we go further based on these works to investigate how the $T\alpha$ and $E\alpha$ correlations affect the power spectra.

To get the power spectra of the CMB anisotropies, we need to expand the temperature and polarization
fields in terms of appropriate spin-weighted spherical harmonic
functions \cite{Zaldarriaga:1996xe}: \bea
T(\bm n)&=& \sum_{lm}a^{T}_{lm}Y_{lm}(\bm n)\nonumber \\
(Q\pm iU) (\bm n)&=& \sum_{lm} a_{\pm 2, lm} \;_{\pm 2}Y_{lm}(\bm n)~.
\eea
The expressions for the expansion coefficients are obtained from the inverse transformations, 
\begin{eqnarray}
a^{T}_{lm}&=&\int d\Omega\; Y_{lm}^{*}(\bm n) T(\bm n)
\nonumber  \\
a_{\pm 2,lm}&=&\int d\Omega \;_{\pm 2}Y_{lm}^{*}(\bm n) (Q\pm iU)(\bm n)~.\label{alm}
\end{eqnarray}
Instead of $a_{2,lm}$ and $a_{-2,lm}$, it is more convenient to use their
linear combinations
\begin{eqnarray}
a^{E}_{lm}=-(a_{2,lm}+a_{-2,lm})/2 \nonumber \\
a^{B}_{lm}=i(a_{2,lm}-a_{-2,lm})/2~.
\label{aeb}
\end{eqnarray}
there are same expansions for the rotated CMB maps where variables are denoted by tildes over them. 
The advantage of $E/B$ decomposition is that it is coordinate-independent and the $E$ and
$B$ modes represent polarization patterns with opposite parities.
Together with the expansion (\ref{alphaex}), the angular power spectra for the auto- and cross-correlations are defined as \beq \langle a^{X'\ast}_{l^\prime
	m^\prime} a^{X}_{lm}\rangle= C^{X'X}_{l} \delta_{l^\prime l}
\delta_{m^\prime m} \eeq with the assumption of statistical isotropy.
Here, $X'$ and $X$ denote the anisotropic rotation angle $\delta \alpha$, the temperature fluctuation $T$ and
the $E$ and $B$ mode polarizations. For the unrotated CMB fields, $C^{TB}_l=C^{EB}_l=0$. As mentioned before, $\delta\alpha$ as scalar perturbation is uncorrelated with primordial $B$ mode, which is sourced by the primordial gravitational waves, so $C^{B\alpha}_l=0$.

Now we move to the discussion of the rotated power spectra. 
Since
\bea
\widetilde{a}_{\pm2, l m} &=& \int d\Omega\; \cdot (\widetilde{Q}\pm i\widetilde{U})({\bm n})\;_{\pm 2}Y^*_{l m}({\bm n}),
\eea
we obtain the following expressions for the rotated $E,B$ fields
\bea
\widetilde{a}^E_{l m}
&=& \frac{1}{2}  \sum_{s, p\geq2, q}e^{is\bar{\alpha}} \Big[ a^E_{pq} +i~ \mathrm{sgn}(s)a^B_{pq} \Big] \int d\Omega\; \cdot e^{is\delta\alpha( {\bm n})}\cdot\;{}_sY_{pq}({\bm n})\;\; {}_sY_{lm}^*({\bm n}),  \nn\\
\widetilde{a}^B_{l m}
&=& \frac{1}{2} \sum_{s, p\geq2, q}  e^{is\bar{\alpha}} \Big[-i ~\mathrm{sgn}(s) a^E_{pq} + a^B_{pq}   \Big]  \int d\Omega\; \cdot e^{is\delta\alpha( {\bm n})}\cdot\;{}_sY_{pq}({\bm n})\;\; {}_sY_{lm}^*({\bm n}).
\label{rotated_harmonic}
\eea
where the subindex $s$ of the spin-weighted spherical harmonic functions ${}_sY_{pq}({\bm n})$ is restricted to be $s=2, -2$ and $\mathrm{sgn}(s)$ is sign function.

The exponent function $e^{is\delta\alpha( {\bm n})}$ has a log-normal distribution, as we will see in the next section this causes the rotated polarization fields to deviate from Gaussian distribution. Actually the non-Gaussian effect has already been noted in ref. \cite{Li:2013vga}, where the authors ignored it by simply assuming that the correlations $T\alpha$ and $E\alpha$ are ignorable. Here we will consider these correlations and in this section we will have a modified version for the CMB power spectra. In the next section, we will calculate the bispecta, and see that the bispectra crucially depend on $T\alpha$ and $E\alpha$ correlations.

First consider the cross power spectra of the temperature with rotated polarizations, for example,   
\bea
\llangle a_{l_1m_1}^{T*} \widetilde{a}^{B}_{l_2m_2} \rrangle &=&     \frac{1}{2} e^{is\bar{\alpha}} \sum_{ s=\pm 2, p\geq2, q}  \int d\Omega\;  \cdot\;{}_sY_{pq}({\bm n})\;\; {}_s{Y}{^*_{l_2 m_2}}({\bm n})\times\nn\\
&&\Big[ \llangle a_{l_1m_1}^{T*} a^B_{pq}  e^{is\delta\alpha( {\bm n})} \rrangle   -i \cdot \mathrm{sgn}(s)  \llangle a_{l_1m_1}^{T*} a^E_{pq}  e^{is\delta\alpha( {\bm n})} \rrangle \Big] ,
\label{TB_power}
\eea
the kernel part the rotated power $TB$ power spectra is a general three point term form $\llangle a^{T}_{l_1m_1}a^{X}_{l_2m_2}e^{is\delta\alpha( {\bm n})}\rrangle, ~X=E,B$. Using the so called Gaussian integration by part formula \cite{vershynin_2018} (see eq.~(\ref{expectation2}) for detail),  this term can be expanded as
\bea
\llangle a^{T*}_{l_1m_1}a^{X}_{pq}e^{is\delta\alpha( {\bm n})}\rrangle =  e^{-2 \llangle \delta\alpha^2( {\bm n})\rrangle  } \bigg[\Big\langle a^{T*}_{l_1m_1}a^{X}_{pq}\Big\rangle - 4\Big\langle a^{T*}_{l_1m_1}\delta\alpha( {\bm n})\Big\rangle \cdot \Big\langle a^{X}_{pq} \delta\alpha( {\bm n})\Big\rangle\bigg],
\label{general_double_corr_expansion}
\eea  
the second term is the modification to previous results in literature by cross correlation of $\delta\alpha$ with CMB. 

Using the expansion (\ref{alphaex}) of the rotation angle,  we get 
\bea
\llangle a_{l_1m_1}^{T*} a^{X}_{pq}  e^{is\delta\alpha( {\bm n})} \rrangle
&=& e^{ -2C^\alpha(0)} \Big[ \llangle a_{l_1m_1}^{T*} a^X_{pq} \rrangle -\nn\\
&&4 \sum_{L_iM_i, i=1}^2\llangle a_{l_1m_1}^{T*} \alpha_{L_1M_1} \rrangle \llangle  a^X_{pq}\alpha_{L_2M_2} \rrangle Y_{L_1M_1}(\bm n) Y_{L_2M_2}(\bm n)\Big], 
\label{TXA}
\eea
where $C^\alpha(0) = \llangle \delta\alpha^2(\bm n)\rrangle $ is the variance of anisotropic rotation angle. Substitute eq.~(\ref{TXA}) into  eq.~(\ref{TB_power}), and use the properties of spin weighted spherical harmonic, we can solve the integration analytically. In calculation, we find the second term of right hand side of eq.~(\ref{TXA}) is eliminated due to the addition theorem of spin weighted spherical harmonics $\sum_{q} \;{}_sY_{pq}({\bm n}) Y_{pq}^*(\bm n) \sim \delta_{0s}=0$, where $s=\pm 2 \neq0$.  Therefore  $C_l^{T\alpha}$ and $C_l^{E\alpha}$ give rise to no contribution to  $\widetilde{C}_l^{TB}$, so as for $\widetilde{C}_l^{TE}$.  After some simplifications, we get exactly the same formula of  $\widetilde{C}_{l}^{TB}, \widetilde{C}_{l}^{TE}$ power spectra  as the result in ref.~\cite{Li:2013vga},
\bea
\tilde{C}^{TE}_l=C_l^{TE}\cos(2\bar{\alpha})e^{-2C^{\alpha}(0)}~,~\tilde{C}^{TB}_l=C_l^{TE}\sin(2\bar{\alpha})e^{-2C^{\alpha}(0)}~.
\eea

We might see why $T\alpha$ and $E\alpha$ correlations have no contribution to the rotated $\tilde{C}_l^{TE}$ and $\tilde{C}_l^{TB}$ spectra in a more
intuitive way. The  spectra $\tilde{C}_l^{TE}$ and $\tilde{C}_l^{TB}$ can be read from the cross correlation function between temperature
and polarization in different direction $\bm n_1, \bm n_2$,   $ \langle T(\bm n_1)(\tilde{Q}_r+i\tilde{U}_r)(\bm n_2)\rangle = e^{2i\bar{\alpha}} \langle T(\bm n_1)e^{2i\delta\alpha(\bm n_2)}(\tilde{Q}_r+i\tilde{U}_r)(\bm n_2)\rangle$~\cite{Li:2013vga, Coulson:1994qw}, here $Q_r, U_r$ are stokes parameters defined with respect to the axes which are parallel and perpendicular to the great arc that connecting the two points. This coordinate choice for stokes parameter makes the correlation function rotational invariant.  For $\bm n_1= \bm n_2$ case, the scalar temperature variable keeps invariant while the spin two stokes parameter varies under rotations around $\bm n_1$. To preserve isotropy, their correlation should vanish. Now expand the rotated correlated function with Eq~.(\ref{expectation2}), we see that the contributions of $T\alpha$
and $E\alpha$ correlations to  $\tilde{C}_l^{TE}$ and $\tilde{C}_l^{TB}$ only come from the combination  $\langle T(\bm n_1) \delta\alpha(\bm n_2)\rangle   \langle \delta\alpha(\bm n_2) ( {Q}_r+i {U}_r)(\bm n_2)\rangle $. Since the the second factor vanishes, $T\alpha$ and $E\alpha$ can not contribute to the rotated 
$\tilde{C}_l^{TE}$ and $\tilde{C}_l^{TB}$ power spectra.

When the two point function contains only polarization modes, the story is different. For example, we calculate the $BB$ correlation as follows,
\bea
&&\llangle \widetilde{a}_{l_1m_1}^{B*} \widetilde{a}^{B}_{l_2m_2} \rrangle \nn\\
&=& \frac{ e^{i\left( s_2  -s_1 \right)\bar{\alpha}} }{4} \sum_{ p_i\geq2, q_i, i=1}^2 \Bigg\{\int d{\Omega_1}   \;{}_{s_1}Y{^*_{p_1q_1}}({\bm n_1})\;\; {}_{s_1}Y{_{l_1 m_1}}({\bm n_1}) \int d{\Omega_2}   \;{}_{s_2}Y{_{p_2q_2}}({\bm n_2})\;\; {}_{s_2}Y{^*_{l_2 m_2}}({\bm n_2}) \nn\\
&\times&\sum_{s_1s_2} \Bigg\langle \bigg[\mathrm{sgn}(s_1s_2)  a^{E*}_{p_1q_1}   a^{E}_{p_2q_2}  +i ~\mathrm{sgn}(s_1) a^{E*}_{p_1q_1}   a^{B}_{p_2q_2} \nn\\
&& \hspace{1cm} -i ~\mathrm{sgn}(s_2) a^{B*}_{p_1q_1}   a^{E}_{p_2q_2}  + a^{B*}_{p_1q_1}   a^{B}_{p_2q_2} \bigg]  e^{i\left[ s_2\delta\alpha( {\bm n_2}) -s_1\delta\alpha( {\bm n_1})\right]}  \Bigg\rangle\Bigg\}, 
\label{BB_power} 
\eea
a general three point term as below needs to be expanded,   here $X_1, X_2 =E, B$, 
\bea
&&\llangle a_{p_1q_1}^{X_1*} a^{X_2}_{p_2q_2}  e^{i\left[ s_2\delta\alpha( {\bm n_2}) -s_1\delta\alpha( {\bm n_1})\right]}\rrangle \nn\\
&=& e^{-4C^\alpha(0) + s_1s_2C^\alpha(\beta)} \Big[ \llangle  a^{X_1*}_{p_1q_1}   a^{X_2}_{p_2q_2} \rrangle -\sum_{L_iM_i, i=1}^2\llangle  a^{X_1*}_{p_1q_1}\alpha_{L_1M_1}\rrangle \llangle  a^{X_2}_{p_2q_2} \alpha_{L_2M_2}\rrangle \nn\\
&& \times \left (s_2\cdot Y_{L_1M_1}( {\bm n_2}) -s_1\cdot Y_{L_1M_1}( {\bm n_1})\right)  \left (s_2\cdot Y_{L_2M_2}( {\bm n_2}) -s_1\cdot Y_{L_2M_2}( {\bm n_1})\right)  \Big],
\label{PPa}
\eea
where $C^\alpha(\beta)$ is the two point correlation function of $\delta\alpha$ over two different directions separated by an angle $\beta$ with $\cos\beta=\bm n_1\cdot \bm n_2$, it relates to the power spectrum $C_l^{\alpha\alpha}$ in the following way, 
\bea
C^\alpha(\beta) = \llangle \delta\alpha(\bm n_1) \delta\alpha(\bm n_2) \rrangle = \sum_L \frac{2L+1}{4\pi} C_L^{\alpha\alpha}P_L(\cos \beta)~,
\label{alpha_correlation}
\eea
here $P_L(\cos \beta)$ is the Legendre polynomial. 

The rest work is substituting eq.~(\ref{PPa}) into eq.~(\ref{BB_power}) and simplifying the integration. The integration over two directions can be simplified as the single integration over the angle between them. Using eqs.~(\ref{wigner2_diagonal}, \ref{wigner3_diagonal}), the explicit expression for rotated $BB$ power spectrum is obtained. In the same way  $EB, EE$ power spectra are calculated. In all, the explicit expressions of rotated polarization power spectra are listed below, 
\bea
\widetilde{C}_{l}^{EE} + \widetilde{C}_{l}^{BB}
&=&  \frac{1}{2}e^{-4 C^\alpha(0)} \int d\cos\beta ~   e^{4 C^\alpha(\beta)}d^{l}_{22}(\beta)\nn\\
&\times&  \Bigg[ \sum_L  (2 L+1)      d^L_{22}(\beta) 
\left(C_{L}^{EE}+C_{L}^{BB}\right) +     W^2_{E\alpha}(\beta)       \Bigg], \nn\\
\widetilde{C}_{l}^{EE} - \widetilde{C}_{l}^{BB}
&=&  \frac{1}{2} e^{-4 C^\alpha(0)} \cos (4 \bar{\alpha} ) \int d\cos\beta ~   e^{-4 C^\alpha(\beta)}d^{l}_{-22}(\beta) \nn\\
&\times&\Big[\sum_L (2 L+1)    d^L_{-22}(\beta)   
\left(C_{L}^{EE} - C_{L}^{BB}\right)  -  W^2_{E\alpha}(\beta)    \Big]  ,\nn\\
\widetilde{C}_{l}^{EB}
&=&  \frac{1}{4} e^{-4 C^\alpha(0)} \sin (4 \bar{\alpha} )\int d\cos\beta ~  e^{-4 C^\alpha(\beta)}d^{l}_{-22}(\beta) \nn\\
&\times&\Bigg[ \sum_L (2 L+1)   d^L_{-22}(\beta)   
\left(  C_{L}^{EE}-C_{L}^{BB}  \right)   - W^2_{E\alpha}(\beta)   \Bigg] ~,\label{rotated_power}
\eea
where $W_{E\alpha}(\beta) $ is defined as
\bea
W_{E\alpha}(\beta) &=& \frac{1}{\sqrt{\pi}}\sum_{L}  \left(2 L+1\right)  d^{L}_{02}(\beta) C_{L}^{E\alpha}~.
\eea
which is a transformation of $C_l^{E\alpha}$ and corresponds to $E\alpha$ correlation in real space. In above calculations, we set $C_l^{B\alpha}=0$. 
We can see that in comparison with the results of ref.~\cite{Li:2013vga}, the corrections to the rotated power spectra only come from $E\alpha$ correlation through the term $W_{E\alpha}(\beta) $, and 
$T\alpha$ correlation has no contribution. This is because the rotation by the Chern-Simons term does not change the temperature map. 

\begin{figure}[hbtp]
	\centering
	\subfigure[]{ 	
	\includegraphics[trim=4.8cm 0 0 0, width=1.08\textwidth]{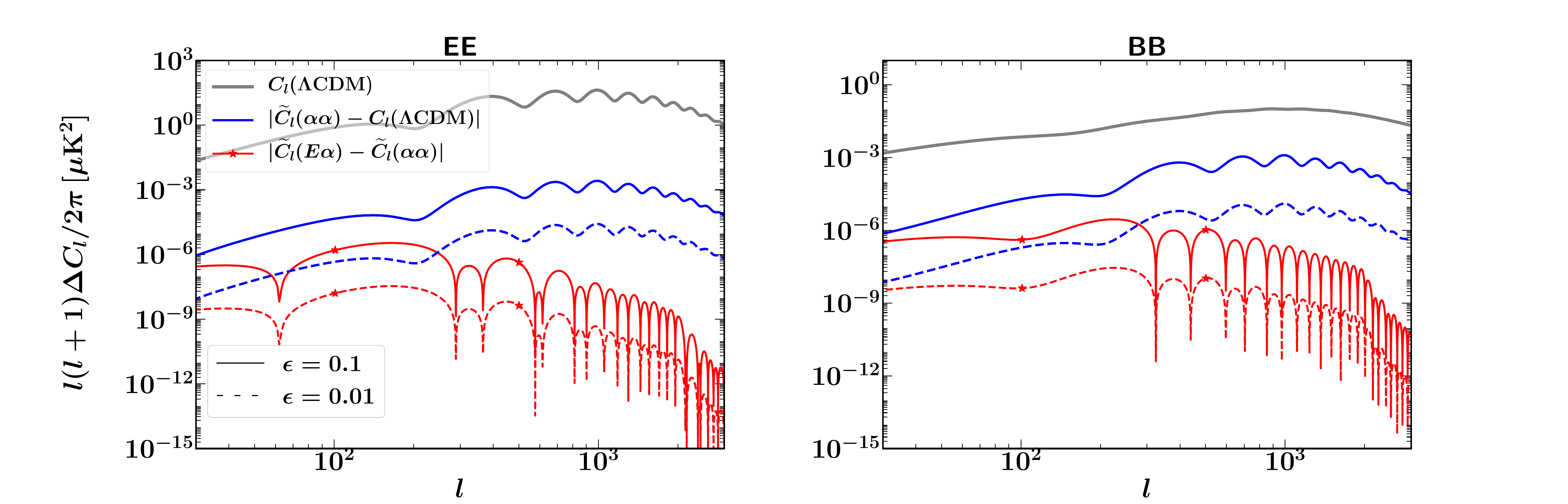}}
	\subfigure[]{ 
	\label{fig_theory_eb}
	\includegraphics[width=3.3in]{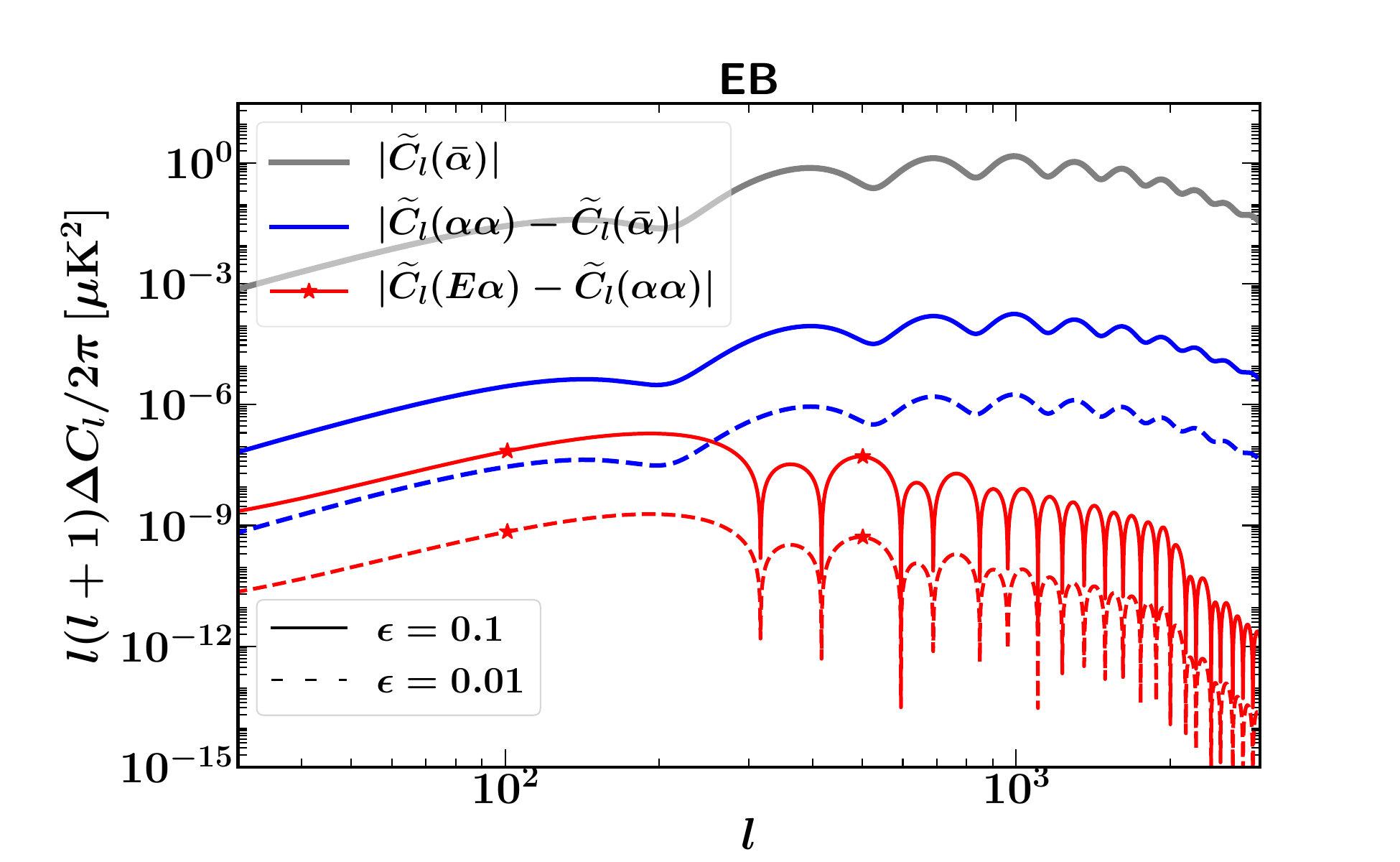}}	
	\caption{Contributions of the $\alpha\alpha$ and $E\alpha$ correlations to the polarized power spectra in Axion-like model. 
		In top two panel, $C_l(\Lambda\mathrm{CDM})$ means the unrotated and lensed CMB power spectra calculated with Planck 2018 best fit model, $\widetilde{C}_l(\alpha\alpha)$ and $\widetilde{C}_l(E\alpha)$ represent the rotated power spectra without and with considering the $E\alpha$ correlations. To be concentrated, we ignore the effect of isotropic rotation angle in top two subfigures by setting $\bar{\alpha}=0$.  In the bottom subfigure we use $\bar{\alpha}=1^\circ$.}
	\label{power_diff}	
\end{figure}

The contribution of $C_l^{\alpha\alpha}$ is encoded in the factor $e^{\pm 4C^\alpha(\beta)}$ (including $e^{-4 C^\alpha(0)}$). From eq.~(\ref{rotated_power}), the leading contribution to the rotated polarization power spectra by auto spectrum of anisotropic rotation angle  are proportional to $C_{L_1}^{\alpha\alpha}C_{L_2}^{EE}$, which is order of $\epsilon^2$.  
While the leading contributions by  $C_l^{E\alpha}$ is proportional to  $C_{L_1}^{E\alpha}C_{L_2}^{E\alpha}$, which is order of $\epsilon^2$ too.  However, in figure~\ref{power_diff}, we see the latter is smaller than the former, and at large multipole their differences increase. This is because the cross power spectrum $C_{L_1}^{E\alpha}$ changes across zero more rapidly, while $C_{L_1}^{\alpha\alpha}C_{L_2}^{EE}$ varies with the multipole moderately which can be seen in figure~\ref{aa_power}. As a result, the summation of former will be larger. In figure~\ref{power_diff}, at about $l \ge 200$, modification to the polarization power spectra by $C_{L_1}^{E\alpha}$ is about two orders of magnitude smaller than the contribution only from $C_{L_1}^{\alpha\alpha}$. Without loss of  accuracy the cross correlation effect can be neglected in such case.  However at small multipoles $l \le 200$, their difference is reduced to only one order. This different  behavior arise from the fact $C_{L_1}^{E\alpha}$ pass zero at about $l \approx 200$. From the expansion of rotated trispectra eq.~(\ref{expectation4}), we can draw similar conclusions. Generally, $C_{L_1}^{E\alpha}$ produces much smaller effect than $C_{L_1}^{\alpha\alpha}$ does so it is more appropriate to use power spectra and trispectra to constrain $C_{L_1}^{\alpha\alpha}$ rather than $C_{L_1}^{E\alpha}$.  To avoid the influence of $C_{L_1}^{E\alpha}$, it is better to use the correlation function at large multipoles. In figure~\ref{power_diff}, we see the curves of $\alpha\alpha$ and $E\alpha$ contributions to the $EE, BB, EB$ power spectra have very similar features, and all the three power spectra are not sensitive with the $E\alpha$ correlation.

\section{Bispectra of rotated CMB fields induced by $T\alpha$ and $E\alpha$ correlations}
\label{section_bispectrum}

In this section, we go further to investigate the CMB bispectra induced by cosmic polarization rotation angle. The procedure of deriving the three point functions is almost the same as that on power spectra except the rotated bispectra are crucially dependent on $T\alpha$ or $E\alpha$ correlations.
Before we get into the complicated  derivations, it's convenient to have a short review on the bispectrum formalism.

Under the assumption of statistical isotropy, the CMB bispectrm should be rotation invariant. As in ref.~\cite{Komatsu:2001rj}, this requires the angular bispectrum $B_{l_1l_2l_3}^{m_1m_2m_3}$ to be proportional to a Wigner 3j factor \cite{Bartolo:2004if}, 
\bea
B_{l_1l_2l_3}^{m_1m_2m_3} &=& \llangle a^X_{l_1m_1} a^Y_{l_2m_2}a^Z_{l_3m_3}\rrangle =  \left( {\begin{array}{ccc}
		l_1 &l_2 & l_3 \\
		m_1 & m_2 & m_3\\
\end{array} } \right)B_{l_1l_2l_3}^{XYZ},
\eea
where $B_{l_1l_2l_3}^{XYZ}$ is called angular averaged bispectrum. Since the spherical harmonic coefficients $a_{l m}^{T/E}$ and $a_{l m}^B$ have the parities $(-1)^{l}$ and $(-1)^{l+1}$  respectively, the parity of  $B_{l_1l_2l_3}^{m_1m_2m_3}$ or $B_{l_1l_2l_3}^{XYZ}$ is $(-1)^{k+l_1+l_2+l_3}$, where $k$ is the number of CMB $B$ fields \cite{Okamoto:2002ik, Kamionkowski:2010rb}. Parity invariance requires all the bispectra should be parity even(hereafter we say parity even means the configuration $k+l_1+l_2+l_3=$even and parity odd for $k+l_1+l_2+l_3=$odd). 

Usually we use the reduced bispectrum instead in order to extract physical information~\cite{Komatsu:2001rj}.  
In the case $l_1+l_2+l_3$=even, the angular averaged bispectrum can be separated into two terms, 
\bea
B_{l_1l_2l_3}^{m_1m_2m_3} &=&\mathcal{G}_{l_1l_2l_3}^{m_1m_2m_3}b_{l_1l_2l_3}, \nonumber\\
\mathcal{G}_{l_1l_2l_3}^{m_1m_2m_3} &=& \int d \Omega~ Y_{l_1 m_1 }(\bm n) Y_{l_2 m_2 }(\bm n) Y_{l_3 m_3 }(\bm n), 
\eea
where $\mathcal{G}_{l_1l_2l_3}^{m_1m_2m_3}$ is named as Gaunt integral which naturally introduces to two basic properties: the triangle inequality $|l_1-l_2|\le l_3 \le |l_1+l_2|$, and the selection rule condition $m_1+m_2+m_3=0$.

The reduced bispectrum with $l_1+l_2+l_3$=odd does not have a unique definition~\cite{Shiraishi:2014roa}. 
Here we follow refs.~\cite{Shiraishi:2014roa, Coulton:2019bnz} and introduce the weight function $h_{l_1l_2l_3}$, therefore the angular averaged bispectrum can be written as
\bea
B_{l_1l_2l_3}^{m_1m_2m_3}& =& b_{l_1l_2l_3} \left( {\begin{array}{ccc}
		l_1 &l_2 & l_3 \\
		m_1 & m_2 & m_3\\
\end{array} } \right)h_{l_1l_2l_3}, \nn\\
h_{l_1l_2l_3}  &=& \frac{1}{6} [1-(-1)^{l_1+l_2+l_3}] \left( I_{l_1l_2l_3}^{11-2} +I_{l_1l_2l_3}^{1-21}+I_{l_1l_2l_3}^{-211} \right)\\
I_{l_1l_2l_3}^{s_1s_2s_3} &=& \sqrt{\frac{(2l_1+1)(2l_2+1)(2l_3+1)}{4\pi}} \left( {\begin{array}{ccc}
		l_1 &l_2 & l_3 \\
		s_1 & s_2 & s_3 \\
\end{array} } \right),\label{wigner_i}
\eea

Finally we obtain a general expression of reduced bispectrum for both cases, which is
\bea
b_{l_1l_2l_3} & =& h^{-1}_{l_1l_2l_3} \left( {\begin{array}{ccc}
		l_1 &l_2 & l_3 \\
		m_1 & m_2 & m_3\\
\end{array} } \right)^{-1}  B_{l_1l_2l_3}^{m_1m_2m_3},\nn\\
h_{l_1l_2l_3} &=&\begin{cases}I_{l_1l_2l_3}^{000}, & l_1+l_2+l_3=\mathrm{even}\\ \frac{1}{6} [1-(-1)^{l_1+l_2+l_3}] \left( I_{l_1l_2l_3}^{11-2} +I_{l_1l_2l_3}^{1-21}+I_{l_1l_2l_3}^{-211} \right), & l_1+l_2+l_3=\mathrm{odd} \end{cases},
\label{geom_factor}
\eea
where $h$ as a generic weight function has different expressions depending on the $l$ domain.

Now let us move on to derive the explicit expressions of rotated bispectra $\langle XYZ \rangle$ where $X, Y, Z$ are harmonic coefficients of rotated CMB fields. For simplicity, we assume all the unrotated CMB temperature and polarization fields are Gaussian random fields and ignore other non-Gaussianity sources from late time evolutions, such as weak lensing and so on. And we also ignore primordial and lensed $B$  mode since their contributions on bispectra are much smaller than that of $E$ polarization and besides $B$ mode adds the computational complexity significantly.
Under these assumptions the rotated CMB harmonic coefficients turn out to be as follows,
\bea
\widetilde{a}^E_{l m}
&=& \widetilde{a}^E_{l m, \mathrm{aniso}} ~\cos{2\bar{\alpha}}  -  \widetilde{a}^B_{l m, \mathrm{aniso}}~ \sin{2\bar{\alpha}}  , \nn\\
\widetilde{a}^B_{l m}
&=&  \widetilde{a}^B_{l m, \mathrm{aniso}} ~\cos{2\bar{\alpha}} +   \widetilde{a}^E_{l m, \mathrm{aniso}} ~\sin{2\bar{\alpha}} , \\
\label{EB_harmonic_scalar}
\widetilde{a}^E_{l m, \mathrm{aniso}}
&=& \frac{1}{2}  \sum_{s, p\geq2, q} a^E_{pq}   \int d\Omega\; \cdot e^{is\delta\alpha( {\bm n})}\cdot\;{}_sY_{pq}({\bm n})\;\; {}_sY_{lm}^*({\bm n}) \nn\\
\widetilde{a}^B_{l m, \mathrm{aniso}}
&=& -\frac{i}{2}  \sum_{s, p\geq2, q} \mathrm{sgn}(s) a^E_{pq}  \int d\Omega\; \cdot e^{is\delta\alpha( {\bm n})}\cdot\;{}_sY_{pq}({\bm n})\;\; {}_sY_{lm}^*({\bm n})~.
\label{aniso_harmonic_scalar}
\eea 

Similar to the case of rotated power spectra, derivations of bispectra eventually result in the computing the ensemble averages of the product of three unrotated CMB harmonic coefficients and one exponent term.
After performing integration by parts as eq.~(\ref{expectation3}), the kernels turn out to be
\bea
\llangle a^{X_1}_{l_1m_1}a^{X_2}_{l_2m_2}a^{X_3}_{l_3m_3}e^{if(\delta\alpha( {\bm n_j}))}\rrangle &=& i e^{-\mathrm{Var}[f(\delta\alpha( {\bm n_j}))]/2 } \bigg[\Big\langle a^{X_1}_{l_1m_1}a^{X_2}_{l_2m_2}a^{X_3}_{l_3m_3} f(\delta\alpha( {\bm n_j}))\Big\rangle \nn\\
&-& \Big\langle a^{X_1}_{l_1m_1}f(\delta\alpha( {\bm n_j}))\Big\rangle \Big\langle a^{X_2}_{l_2m_2} f(\delta\alpha( {\bm n_j}))\Big\rangle  \Big\langle  a^{X_3}_{l_3m_3} f(\delta\alpha( {\bm n_j}))\Big\rangle \bigg]~,\nn\\
\label{general_tri_corr_expansion} 
\eea
where $X_1, X_2 = T, E, B$, $X_3 = E, B$, the exponent argument is a linear combination of anisotropic rotation angles $f(\delta\alpha( {\bm n_j}))=\sum_j s_j\delta\alpha( {\bm n_j})$. The variance $\mathrm{Var}\left[f(\delta\alpha( {\bm n_j}))\right]$ reads , 
\bea
\mathrm{Var}\left[ \sum_{i} s_i\delta\alpha(\bm n_i) \right] 
=    \sum_{i, j} s_i s_j  \llangle \delta\alpha(\bm n_i) \delta\alpha(\bm n_j) \rrangle =  \sum_{i,j,L} \frac{2L+1}{4\pi} s_i s_j C_L^{\alpha\alpha}P_L\left(\cos (\bm{n_i\cdot n_j})\right),  
\eea

It is convenient to categorize those bispectra according to the number of polarization fields, namely $TTP$, $TPP$ and $PPP$, here $P$ means polarization.
For the simplest case $TTP$, we have
\bea
\llangle a_{l_1m_1}^T {a}^T_{l_2m_2}  \widetilde{a}^E_{l_3m_3} \rrangle 
&=& \frac{1}{2} \sum_{s } ~ \mathcal{I}_1(s,  l_{123}), \nn\\
\llangle a_{l_1m_1}^T {a}^T_{l_2m_2}  \widetilde{a}^B_{l_3m_3} \rrangle  
&=&-\frac{i}{2} \sum_{s }  \mathrm{sgn}(s)  \mathcal{I}_1(s ,  l_{123}),
\label{ttp_definition} 
\eea
we introduce $\mathcal{I}_1$ that represents the kernel integration in the case of $TTP$, defined as,
\bea
\mathcal{I}_1(s, l_{123})&= & e^{is\bar{\alpha}}  \sum_{ p\geq2, q}  \int d\Omega\;  \;{}_sY_{pq}({\bm n})\;\; _s{Y}{^*_{l_3m_3}}({\bm n}) \llangle a_{l_1m_1}^{T} a_{l_2m_2}^{T} a_{pq}^{E} e^{ i s \delta\alpha( {\bm n})}\rrangle \nn\\
&=&  i~s~ e^{   is\bar{\alpha}  -2C^\alpha(0) }     C_{l_1}^{TE} C_{l_2}^{T\alpha}  I^{0-ss}_{l_2l_1l_3} \left( {\begin{array}{ccc}
		l_1 & l_2 & l_3 \\
		m_1 & m_2 & m_3\\
\end{array} } \right) +( l_1, m_1 \leftrightarrow l_2, m_2)~,
\eea
where $l_{123}$ is abbreviation for $l_1, l_2, l_3$. In the simplification, we used the formula for the expectation of product in eq.~(\ref{general_tri_corr_expansion}), and then expanded $\delta\alpha(\bm n)$ in terms of spherical harmonics.

Substitute the expression of $\mathcal{I}_1(s,l_{123} )$ into eq.~(\ref{ttp_definition}) and combine with the definition of reduced bispectrum, we obtain 
\bea
\widetilde{b}^{TTE}_{l_1l_2l_3} 
&=&  2 h^{-1}_{l_1l_2l_3} I^{0-22}_{l_2l_1l_3}   e^{ -2C^\alpha(0) }   C_{l_1}^{TE} C_{l_2}^{T\alpha}  \Big[i\mathcal{O}_{l_{\mathrm{sum}}} \cos(2\bar{\alpha})-\mathcal{E}_{l_{\mathrm{sum}}}   \sin(2\bar{\alpha})\Big] + ( l_1 \leftrightarrow l_2),\nn\\
\widetilde{b}^{TTB}_{l_1l_2l_3}  
&=& 2 h^{-1}_{l_1l_2l_3} I^{0-22}_{l_2l_1l_3}   e^{ -2C^\alpha(0) }  C_{l_1}^{TE} C_{l_2}^{T\alpha}   \Big[\mathcal{E}_{l_{\mathrm{sum}}}     \cos(2\bar{\alpha}) +i\mathcal{O}_{l_{\mathrm{sum}}}   \sin(2\bar{\alpha})  \Big] + ( l_1 \leftrightarrow l_2)~,
\label{reduced_TTP}
\eea
where $\mathcal{E}_{l_{\mathrm{sum}}}, \mathcal{O}_{l_{\mathrm{sum}}}$ are even/odd symbols which are defined as
\bea
\mathcal{E}_{l_{\mathrm{sum}}} &=& \left[1+(-1)^{l_{\mathrm{sum}}}\right]\Big/2, ~~ \mathcal{O}_{l_{\mathrm{sum}}} = \left[1-(-1)^{l_{\mathrm{sum}}}\right]\Big/2,~~l_{\mathrm{sum}} = l_1+l_2+l_3~. 
\eea
In eq. (\ref{reduced_TTP}), the reduced $TTP$ bispectra contain the opposite parity terms, which means parity violation. This violation has two origins. The first one comes from the parity mixing brought by the isotropic rotation angle $\bar{\alpha}$, which depends on the background evolution. The second one comes from the anisotropic rotation angle. This can be seen as the limit the rotation angle has no background part, $\bar{\alpha} = 0$.  At this limit only the odd-$TTE$ (hereafter the prefix ``odd-'' means $l_{\mathrm{sum}}=\mathrm{odd}$) and even-$TTB$ (hereafter the prefix ``even-''  means $l_{\mathrm{sum}}=\mathrm{even}$) components left.  Both of them are parity odd bispectra \cite{Okamoto:2002ik} and break the parity conservation. 
Furthermore, we can see that the $TTP$ bispectra are proportional to $C_l^{T\alpha}$. That is, without $T\alpha$ correlation, both the $TTE$ and $TTB$ bispectra for the rotated CMB polarization fields vanish.

\begin{figure}[t]
	\centering
	\includegraphics[trim=4.8cm 0 0 0, width=1.08\textwidth]{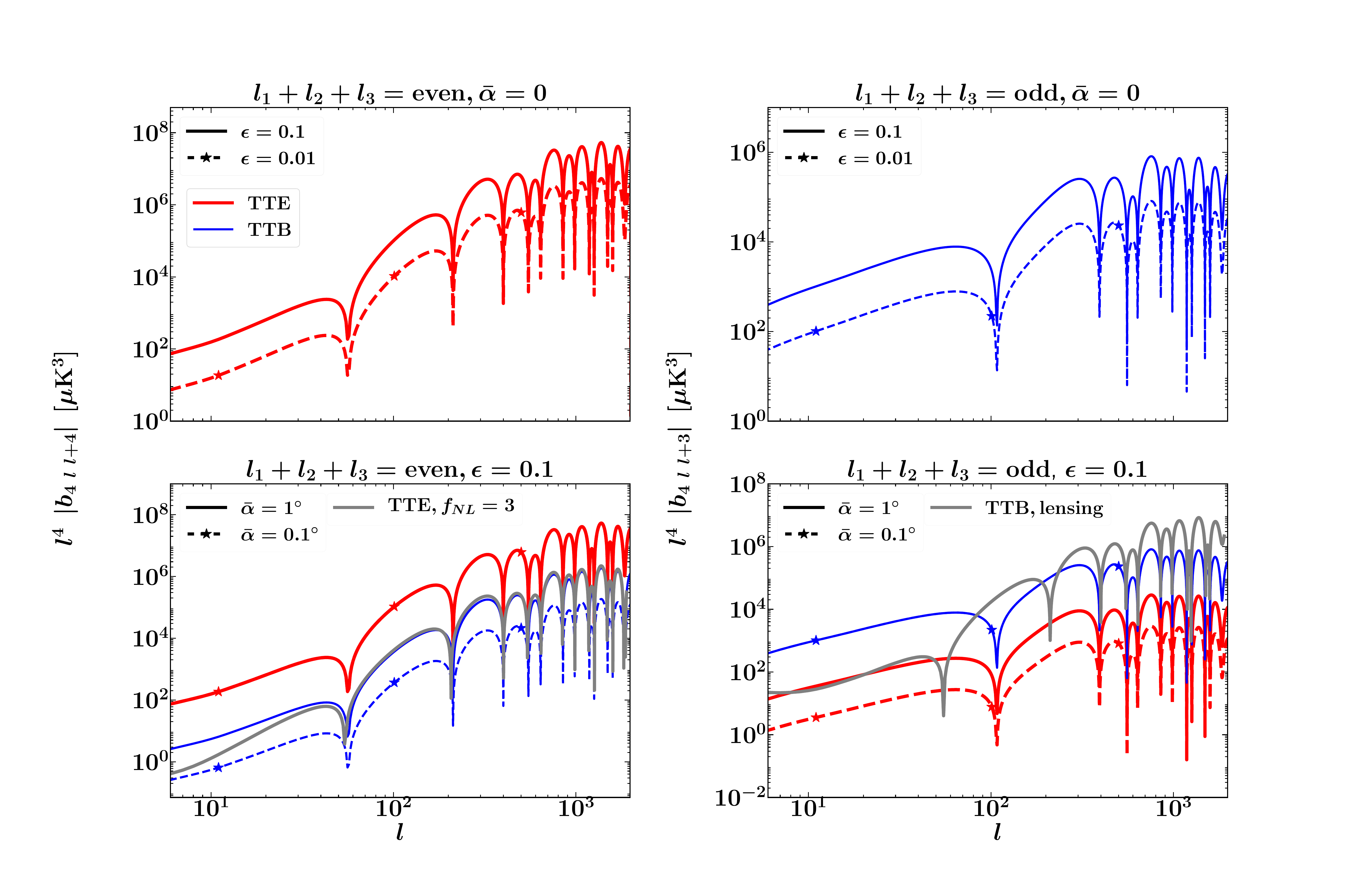}
	\caption{Absolute values of rotated reduced bispectra  $b^{TTP}_{4l_2l_3}$  as functions of $l_2$ for various parameter and parities.  The input $C_l^{T\alpha}$ and $C_l^{E\alpha}$ are based on figure~\ref{aa_power}. Multipoles are set to be $l_3=l_2+4$ for even domain and $l_3=l_2+3$ for odd domain. Note for bispectrum in odd domain, the imaginary number is omitted. Top panels: $\bar{\alpha} = 0$, the solid and dashed curves correspond to parameter $\epsilon=0.1, 0.01$. Bottom panels: $\epsilon=0.1$, the solid and dashed curves correspond to $\bar{\alpha}=1^\circ, 0.1^\circ$. In bottom left/right panels, the primordial bispectra even-$TTE $ with the local non-linear parameter $f_{NL}=3$, and  weak lensing induced odd-$TTB$ are plotted for comparison.}
	\label{bis_TTP}	
\end{figure}

In figure~\ref{bis_TTP} we plot the rotation angle induced $TTP$ bispectra with different choices of parameters, and the primordial $TTE$ and lensed $TTB$ for comparisons.  
We can see that parity-violating bispectra even-$TTB$ and odd-$TTE$, which remains zero in the case of primordial scalar non-Gaussianity and weak lensing, will be generated if there exists non-zero cross-correlation between CMB temperature field and anisotropic rotation angle. In addition they are almost unaffected by the isotropic rotation angle.
These features make even-$TTB$ and odd-$TTE$ bispectra good estimators for probing anisotropic rotation angles and the cross-correlation $C_l^{T\alpha}$.
Furthermore, in order to estimate $\epsilon$, since the reduced bispectra is proportional to $\epsilon$ because of $C^{T\alpha}_{l}$ in eq.~(\ref{reduced_TTP}), and  rotated CMB power spectra is proportional to $\epsilon^2$ to its leading order, odd-$TTE$ and even-$TTB$ also could potentially perform better than the power spectra.
For the parity conserved bispectra sourced by a nonzero isotropic rotation angle,  even-$TTE$ and odd-$TTB$ will potentially affect probing other non-Gaussianity sources similar to how rotation angle does to primordial gravitational waves' observations.

For the cases when the three point functions contain two polarization fields $TPP$, the bispectra share similar forms
\bea
\llangle a_{l_1m_1}^T \widetilde{a}^E_{l_2m_2}  \widetilde{a}^E_{l_3m_3} \rrangle 
&=& \frac{1}{4} \sum_{s_1s_2} ~ \mathcal{I}_2(s_{12},  l_{123}) \nn\\
\llangle a_{l_1m_1}^T \widetilde{a}^E_{l_2m_2}  \widetilde{a}^B_{l_3m_3} \rrangle  
&=&-\frac{i}{4} \sum_{s_1s_2}  \mathrm{sgn}(s_2)  \mathcal{I}_2(s_{12},  l_{123}) \nn\\
\llangle a_{l_1m_1}^T \widetilde{a}^B_{l_2m_2}  \widetilde{a}^B_{l_3m_3} \rrangle  
&=& -\frac{1}{4} \sum_{s_1s_2}\mathrm{sgn}(s_1s_2)    \mathcal{I}_2(s_{12},  l_{123}),
\label{TPP_definition}
\eea
where $\mathcal{I}_2$ is the integration over two directions $\bm n_1, \bm n_2$,  
\bea
\mathcal{I}_2(s_{12},  l_{123})
= e^{i\sum_{j=1}^2s_j\bar{\alpha}} \sum_{p_i,q_i, i=1}^2  
&&\int d{\Omega_1}  \;{}_{s_1}Y{_{p_1q_1}}({\bm n_1})\;\; {}_{s_1}Y{^*_{l_2m_2}}({\bm n_1})\nn\\
&\times&\int d{\Omega_2}  \;{}_{s_2}Y{_{p_2q_2}}({\bm n_2})\;\; {}_{s_2}Y{^*_{l_3m_3}}({\bm n_2}) \nn\\
&\times& \llangle a_{l_1m_1}^{T} a_{p_1q_1}^{E} a_{p_2q_2}^{E} e^{i\sum_{j=1}^2s_j\delta\alpha( {\bm n_j})} \rrangle~,
\label{TPP_type_int}
\eea
here $s_{12}$ is short for $s_1, s_2$.
Expanding the ensemble average part and then simplify the integration using eqs.~(\ref{wigner2_diagonal}, \ref{wigner3_diagonal}), we obtain
\begin{align}
\mathcal{I}_2(s_{12},  l_{123})  
&= i e^{i\sum_{j=1}^2s_j\bar{\alpha}} e^{-4C^\alpha(0) }  \left( {\begin{array}{ccc}
	l_1 &l_2 & l_3 \\
	m_1 & m_2 & m_3\\
	\end{array} } \right) \int d\cos\beta ~ e^{- s_1 s_2 C^\alpha (\beta) } \Bigg[ s_1C_{l_1}^{{TE}} I^{0s_1-s_1}_{l_3l_2l_1}d^{l_3}_{0s_2}(\beta)      \nn\\
&\times \sum_{L}    \frac{2L+1}{2}C_{L}^{E\alpha}     d^{L}_{0s_2}(\beta) + s_1C_{l_1}^{T\alpha} I_{l_1l_2l_3}^{0-s_1s_1}d^{l_3}_{-s_2s_1}(\beta)  \sum_{L}   \frac{2L+1}{2}  C_{L}^{EE}      d^{L}_{-s_2s_1}(\beta)     \nn\\
& -   \frac{1}{2\pi} s_2 C_{l_1}^{T\alpha} I^{0-s_1s_1}_{l_1l_2l_3} d^{l_3}_{-s_2s_1}(\beta)\sum_{L_1} (2L_1+1)C_{L_1}^{E\alpha}d^{L_1}_{0s_1}(\beta)\sum_{L_2}(2L_2+1)  C_{L_2}^{E\alpha}     d^{L_2}_{0s_2}(\beta)\Bigg]\nn\\
& + (\{s_1l_2m_2\} \leftrightarrow \{s_2l_3m_3\} )~,
\label{TBB_int_result}
\end{align}
where the permutation means $s_1 \leftrightarrow s_2, l_2 \leftrightarrow  l_3,  m_2 \leftrightarrow  m_3$. Substitute it into eq.~(\ref{TPP_definition}), we obtain the reduced  bispectrum,		
\bea
\widetilde{b}^{TEE}_{l_1l_2l_3} &=& h^{-1}_{l_1l_2l_3}   e^{- 4C^\alpha (0)} \int _0^\pi d\cos\beta ~e^{-4C^\alpha (\beta)} \Big\{\mathcal{E}_{l_{\mathrm{sum}}}\sin(4\bar{\alpha})\Big[-U^{\rmnum{1}}_{l_{123}}-U^{\rmnum{2}}_{l_{123}}+U^{\rmnum{4}}_{l_{123}}  \Big] \nn\\
&&+ i\mathcal{O}_{l_{\mathrm{sum}}}\Big[ -\cos(4\bar{\alpha})\left(  U^{\rmnum{1}}_{l_{123}} -  U^{\rmnum{2}}_{l_{123}}+U^{\rmnum{4}}_{l_{123}}\right) - e^{8C^\alpha (\beta)} U^{\rmnum{1}}_{l_{123}} +U^{\rmnum{3}}_{l_{123}}+U^{\rmnum{5}}_{l_{123}} \Big]  \Big\} \nn\\
&& +(l_2 \leftrightarrow l_3).
\label{bispectra_tee}
\eea
The auxiliary functions $U_{l_{123}}$ are defined as
\bea
U^{\rmnum{1}}_{l_{123}} &=& d^{l_3}_{02}(\beta) I^{0-22}_{l_3l_2l_1}   C_{l_1}^{TE}  \sum_{L}  \frac{2L+1}{2} d^{L}_{02}(\beta)C_{L}^{E\alpha}, \nn\\
U^{\rmnum{2}}_{l_{123}} &=& d^{l_3}_{-22}(\beta) I_{l_1l_2l_3}^{0-22}     C_{l_1}^{T\alpha}    \sum_{L} \frac{2L+1}{2}d^{L}_{-22}(\beta) C_{L}^{EE},\nn\\
U^{\rmnum{3}}_{l_{123}} &=& e^{8C^\alpha (\beta)}  d^{l_3}_{22}(\beta)   I_{l_1l_2l_3}^{0-22}     C_{l_1}^{T\alpha}    \sum_{L} \frac{2L+1}{2}d^{L}_{22}(\beta)C_{L}^{EE}, \nn\\
U^{\rmnum{4}}_{l_{123}} &=& \frac{d^{l_3}_{-22}(\beta)}{2\pi}  I^{0-22}_{l_1l_2l_3}   C_{l_1}^{T\alpha} \left( \sum_{L_1}      (2L_1+1)d^{L_1}_{02}(\beta)C_{L_1}^{E\alpha}\right)^2 , \nn\\
U^{\rmnum{5}}_{l_{123}} &=&\frac{ e^{8C^\alpha (\beta)}d^{l_3}_{22}(\beta)}{2\pi}      I^{0-22}_{l_1l_2l_3}   C_{l_1}^{T\alpha} \left(   \sum_{L_1}      (2L_1+1)d^{L_1}_{02}(\beta)C_{L_1}^{E\alpha}\right)^2.
\label{TPP_auxiliary}
\eea
the subscripts $l_{123}$ represents sequence of $l_1l_2l_3$.
Similarly, the expressions for $TEB$ and  $TBB$ are obtained, 
\bea
\widetilde{b}^{TEB}_{l_1l_2l_3} &=& h^{-1}_{l_1l_2l_3}   e^{- 4C^\alpha (0)} \int _0^\pi d\cos\beta ~ e^{-4C^\alpha (\beta)}
\Big\{\mathcal{E}_{l_{\mathrm{sum}}}\Big[  e^{8C^\alpha (\beta)}\left(  U^{\rmnum{1}}_{l_{132}}-U^{\rmnum{1}}_{l_{123}}\right)  \nn\\
&&+  \cos(4\bar{\alpha})\big(  U^{\rmnum{1}}_{l_{123}} + U^{\rmnum{1}}_{l_{132}}  +  U^{\rmnum{2}}_{l_{123}} +   U^{\rmnum{2}}_{l_{132}}   -    U^{\rmnum{4}}_{l_{123}}+  U^{\rmnum{4}}_{l_{132}}   \big) +   U^{\rmnum{3}}_{l_{132}}-U^{\rmnum{3}}_{l_{123}}    +    U^{\rmnum{5}}_{l_{132}}         -     U^{\rmnum{5}}_{l_{123}}     \Big] \nn\\
&&+ i\mathcal{O}_{l_{\mathrm{sum}}}\sin(4\bar{\alpha})\Big[      U^{\rmnum{1}}_{l_{132}} -     U^{\rmnum{1}}_{l_{123}}  +     U^{\rmnum{2}}_{l_{123}}-   U^{\rmnum{2}}_{l_{132}}  +    U^{\rmnum{4}}_{l_{132}} -    U^{\rmnum{4}}_{l_{123}}     \Big] \Big\}, 
\label{TEB_bispectra}\\
\widetilde{b}^{TBB}_{l_1l_2l_3} &=& h^{-1}_{l_1l_2l_3}   e^{- 4C^\alpha (0)} \int _0^\pi d\cos\beta ~ e^{-4C^\alpha (\beta)}
\Big\{\mathcal{E}_{l_{\mathrm{sum}}}\sin(4\bar{\alpha})\big(U^{\rmnum{1}}_{l_{123}}  + U^{\rmnum{2}}_{l_{123}}-U^{\rmnum{4}}_{l_{123}}   \big)\nn\\
&&+ i\mathcal{O}_{l_{\mathrm{sum}}}\Big[ \cos(4\bar{\alpha})\left(  U^{\rmnum{1}}_{l_{123}} -   U^{\rmnum{2}}_{l_{123}} +   U^{\rmnum{4}}_{l_{123}} \right) - e^{8C^\alpha (\beta)} U^{\rmnum{1}}_{l_{123}}  + U^{\rmnum{3}}_{l_{123}} +U^{\rmnum{5}}_{l_{123}} \Big]\Big\} \nn\\
&&+(l_2 \leftrightarrow l_3).
\label{TBB_bispectra}
\eea
We can see that all the rotated $TPP$ bispectra will vanish if there are no $T\alpha$ or $E\alpha$ correlations. Furthermore, both the parity-odd and parity-even terms are produced.

Under the small rotation angle approximation,
\bea
e^{-4C^\alpha(\beta)} \sim 1, ~~\sin\bar{\alpha}\sim \bar{\alpha}, ~~ \cos \bar{\alpha} \sim 1 ~,
\label{approximate_condition}
\eea
we can get a rough evaluation on the amount of reduced bispectra as follows,
\bea
&\widetilde{b}^{TEE}(\mathrm{even}) \sim ( {\bar{\alpha}}C_l^{T\alpha}, {\bar{\alpha}} C_l^{E\alpha}) , ~~ & \widetilde{b}^{TEE}(\mathrm{odd}) \sim  (  C_l^{T\alpha},   C_l^{E\alpha}), \nn\\
&\widetilde{b}^{TEB}(\mathrm{even}) \sim (  C_l^{T\alpha},   C_l^{E\alpha}), ~~ & \widetilde{b}^{TEB}(\mathrm{odd}) \sim ( {\bar{\alpha}}C_l^{T\alpha}, {\bar{\alpha}} C_l^{E\alpha}), \nn\\
&\widetilde{b}^{TBB}(\mathrm{even}) \sim  ( \bar{\alpha}C_l^{T\alpha}, {\bar{\alpha}} C_l^{E\alpha}), ~ ~ &\widetilde{b}^{TBB}(\mathrm{odd}) \sim (\bar{\alpha}^2 C_l^{T\alpha},  \bar{\alpha}^2 C_l^{E\alpha}, C_l^{T\alpha} C_{L_1}^{E\alpha} C_{L_2}^{E\alpha})~,
\label{TPP_approximation}
\eea
where each approximate equation means the bispectra are at the same order with the terms within the parenthesis. 
See eq.~(\ref{approximate_tpp})) for complete expressions. Again here the word ``even''/``odd'' in the parenthesis means $l_1+l_2+l_3 = $ even/odd.

\begin{figure}[t]
	\centering
	\includegraphics[trim=4.8cm 0 0 0, width=1.08\textwidth]{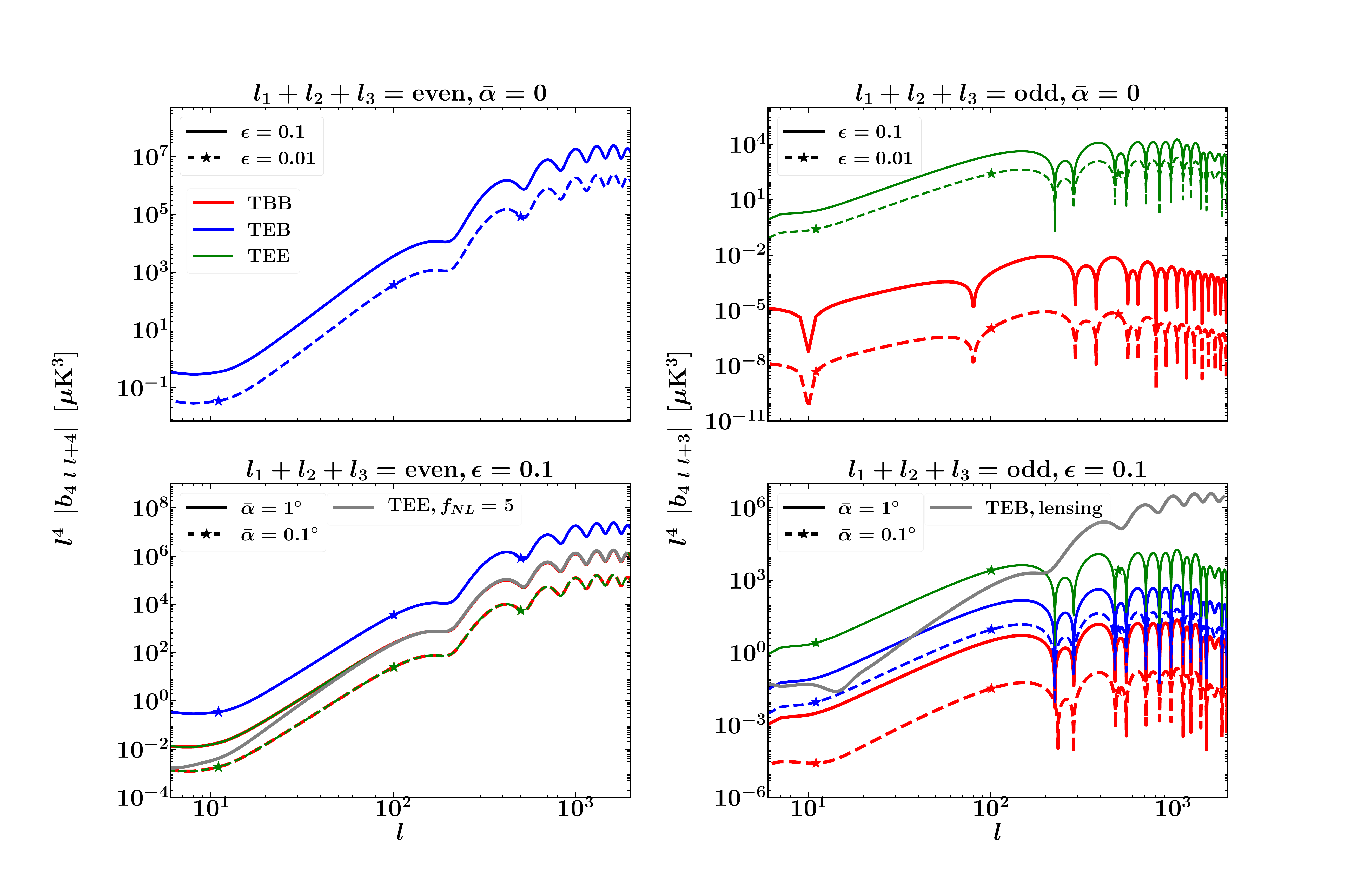}
	\caption{ Absolute values of rotated reduced bispectra  $b^{TPP}_{4l_2l_3}$ as functions of $l_2$ for various parameter and parities.  Configuration of multipole and parameters settings are as same as that in figure~\ref{bis_TTP}. For comparison, the primordial bispectra $ TEE  $ with  $f_{NL}=5$, and weak lensing induced $TEB$ are plotted.}
	\label{bis_TPP}	
\end{figure}

We plot polarization rotation angle induced $TPP$ type bispectra in figure~\ref{bis_TPP}, as well as bispectra introduced by primordial non-Gaussianity and weak lensing for comparison.
Similar to the case of $TTP$, even-$TEB$, odd-$TEE$ and odd-$TBB$ will be generated due to the anisotropic rotation angle.
Combining with the isotropic rotation angle, parity conserved bispectra even-$TEE$, even-$TBB$, odd-$TEB$ will also be created through converting $E$ and $B$ fields with each other. Interestingly the even-$TEE$ and even-$TBB$ are equal since they are both derived by converting even-$TEB$ through isotropic rotation. 
$TPP$ bispectra could still be a set of good estimators for measuring $C_{l}^{T\alpha}$, $C_{l}^{E\alpha}$ and $\epsilon$, since at leading order even-$TEB$ and odd-$TEE$ are linear combinations of $\delta\alpha$ involved cross-correlation spectra, and thus both of them are proportional to $\epsilon$.

The rest cases in which the three point correlation functions contain three rotated polarization fields $PPP$ are expected to have the most complicated forms. In a similar way, we start from the definitions of reduced bispectra, 
\bea
\llangle\widetilde{a}_{l_1m_1}^E \widetilde{a}^E_{l_2m_2}  \widetilde{a}^E_{l_3m_3} \rrangle 
&=& \frac{1}{8} \sum_{s_1s_2s_3} \mathcal{I}_3(s_{123}, l_{123}),\nn\\
\llangle\widetilde{a}_{l_1m_1}^E \widetilde{a}^E_{l_2m_2}  \widetilde{a}^B_{l_3m_3} \rrangle 
&=& -\frac{i}{8} \sum_{s_1s_2s_3} \mathrm{sgn}(s_3)  \mathcal{I}_3(s_{123}, l_{123}),  \nn\\
\llangle\widetilde{a}_{l_1m_1}^E \widetilde{a}^B_{l_2m_2}  \widetilde{a}^B_{l_3m_3} \rrangle 
&=& -\frac{1}{8} \sum_{s_1s_2s_3} \mathrm{sgn}(s_2s_3)  \mathcal{I}_3(s_{123}, l_{123}), \nn\\
\llangle\widetilde{a}_{l_1m_1}^B \widetilde{a}^B_{l_2m_2}  \widetilde{a}^B_{l_3m_3} \rrangle 
&=& \frac{i}{8} \sum_{s_1s_2s_3} \mathrm{sgn}(s_1s_2s_3)  \mathcal{I}_3(s_{123}, l_{123}), 
\label{PPP_definition}
\eea
where $\mathcal{I}_3$ is the integration over three directions $\bm n_1, \bm n_2, \bm n_3$, reads
\bea
\mathcal{I}_3(s_{123},  l_{123})
&=&  e^{i\sum_{j=1}^3 s_j\bar{\alpha}} \sum_{p_i,q_i, i=1}^3  \int d{\Omega_1}  \;{}_{s_1}Y{_{p_1q_1}}({\bm n_1})\;\; {}_{s_1}Y{^*_{l_1m_1}}({\bm n_1})\int d{\Omega_2}  \;{}_{s_2}Y{_{p_2q_2}}({\bm n_2})\;\; {}_{s_2}Y{^*_{l_2m_2}}({\bm n_2}) \nn\\
&& \times \int d{\Omega_3}  \;{}_{s_3}Y{_{p_3q_3}}({\bm n_3})\;\; {}_{s_3}Y{^*_{l_3m_3}}({\bm n_3})  \llangle a_{p_1q_1}^{E}  a_{p_2q_2}^{E} a_{p_3q_3}^{E} e^{i\sum_{j=1}^3 s_j\delta\alpha( {\bm n_j})} \rrangle~.
\label{TBB_int}
\eea

We only consider approximations up to first order of $C_l^{\alpha\alpha}$ in order to simplify the numerical calculation. First expand the expectation then truncate the exponential part to first order,
\bea
e^{- \sum_{(i,j)}^3 s_i s_j C^\alpha (\bm {n_i\cdot n_j})} & \approx & 1 -  \sum_{LM} s_1 s_2 C_{L}^{\alpha\alpha} Y^*_{LM} (\bm n_1) Y_{LM}(\bm n_2) \nn\\
&&-  \sum_{LM} s_2 s_3 C_{L}^{\alpha\alpha} Y^*_{LM} (\bm n_2) Y_{LM}(\bm n_3) \nn\\
&&-  \sum_{LM} s_1 s_3 C_{L}^{\alpha\alpha} Y^*_{LM} (\bm n_1) Y_{LM}(\bm n_3)~.
\eea
Substitute this expansion into expansion of eq.~(\ref{TBB_int}), the integral can be  analytically done using the properties of spin weighted spherical harmonics and Wigner symbols present in appendices \ref{spin-harmonic_property}, \ref{wigner_property}.
After some simplification, the zeroth and first orders of $\mathcal{I}_3$ are
\bea
&&\mathcal{I}_3(s_{123},  l_{123})(0)
=     i e^{i\sum_{j=1}^3 s_j\bar{\alpha}} e^{-6C^\alpha(0) } \left( {\begin{array}{ccc}
	l_1 & l_2 &  l_3 \\
	m_1 & m_2 &  m_3 \\
	\end{array} } \right)\Bigg\{ {s_1}   I_{l_2l_1l_3}^{-s_1s_10}  C_{l_3}^{{E\alpha}}C_{l_2}^{EE}\nn\\
&& -\frac{1}{6}\Big[ s_2s_3s_1 \sum_{L_i, i=1}^3C_{L_1}^{E\alpha} C_{L_2}^{E\alpha}  C_{L_3}^{E\alpha} (-1)^{L_3+L_2+L_1+l_1+l_2+l_3} I_{L_3l_1L_1}^{-s_1s_10} I_{L_2l_2L_3}^{-s_2s_20}I_{L_1l_3L_2}^{-s_3s_30}  \left\{ {\begin{array}{ccc}
	l_1 & l_2 & l_3 \\
	L_2 & L_1 & L_3 \\
	\end{array} }\right\} + (2 \leftrightarrow 3)\Big] \nn\\
&&     - \frac{4s_1}{(2l_1+1) }  C_{l_3}^{E\alpha} I_{l_2l_3l_1}^{s_20-s_2}\sum_{L_i, i=1}^2 (-1)^{l_1 +l_2+l_3 }I_{L_1l_1L_2}^{-s_1s_10}  I_{L_1L_2l_1}^{0-s_2s_2}  C_{L_1}^{E\alpha}C_{L_2}^{E\alpha}\Bigg\} + 5(\mathrm{~perms}),
\label{zero_PPP_int} 
\eea
\bea
&&\mathcal{I}_3(s_{123},  l_{123})(1)
=  ie^{i\sum_{j=1}^3 s_j\bar{\alpha}} e^{-6C^\alpha(0) }
    \left( {\begin{array}{ccc}
	l_1 & l_2 &  l_3 \\
	m_1 & m_2 &  m_3 \\
	\end{array} } \right)\Bigg\{\nn\\
&& -  s_1 s_2 s_3  \sum_{L_i, i=1}^3 I_{L_3l_1L_1}^{-s_1s_10}  I_{L_3l_2L_2}^{-s_2s_20}I_{L_1l_3L_2}^{-s_3s_30}  C_{L_3}^{EE} C_{L_2}^{\alpha\alpha}    C_{L_1}^{{E\alpha}}  (-1)^{L_1+l_1+l_3} \left\{ {\begin{array}{ccc}
	l_1 & l_2 & l_3  \\
	L_2 & L_1 & L_3  \\
	\end{array} } \right\}\nn\\
&& - \frac{4 s_2 }{(2l_2+1)}     C_{l_3}^{{E\alpha}}I^{0-s_1s_1}_{l_3l_1l_2} \sum_{L_i, i=1}^2 (-1)^{L_1+L_2+l_2}    I_{L_2l_2L_1}^{-s_2s_20}   I^{0-s_1s_1}_{L_1L_2l_2} C_{L_1}^{\alpha\alpha}C_{L_2}^{EE}  \nn\\ 
&& -    \frac{4s_3}{(2l_3+1)}  C_{l_2}^{EE} I_{l_2l_1l_3}^{-s_1s_10} \sum_{L_i, i=1}^2   I_{L_2l_3L_1}^{-s_3s_30}I^{000}_{l_3L_2L_1}    C_{L_1}^{\alpha\alpha}    C_{L_2}^{{E\alpha}} \Bigg\}  + 5(\mathrm{~perms})~,
\label{first_PPP_int} 
\eea
where the permutation is for both $s$ and $l, m$.
Note in calculation, terms with more than three orders of $\delta\alpha$, or say $\epsilon$, such as $ \sum_{L_1L_2L_3L_4}C_{L_1}^{E\alpha}C_{L_2}^{E\alpha}C_{L_3}^{E\alpha}C_{L_4}^{\alpha\alpha}$ are omitted.

Substitute eqs.~(\ref{zero_PPP_int}, \ref{first_PPP_int}) into eq.~(\ref{PPP_definition}), we obtain the reduced bispectra
\bea
\widetilde{b}^{EEE}_{l_1l_2l_3}
&=& 2\mathcal{E}_{l_{\mathrm{sum}}}h^{-1}_{l_1l_2l_3}\sin(2\bar{\alpha})\cos^2(2\bar{\alpha})\Big[ -V^0_{l_{123}}+  V^{\rmnum{1}}_{l_{123}}  +  V^{\rmnum{3}}_{l_{123}}+    V^{\rmnum{5}}_{l_{123}}\Big]\nn\\
&&+ i\mathcal{O}_{l_{\mathrm{sum}}} h^{-1}_{l_1l_2l_3}\cos(2\bar{\alpha})  \Big[2\cos^2(2\bar{\alpha})V^0_{l_{123}} +  \cos(4\bar{\alpha})\left(  V^{\rmnum{1}}_{l_{123}} - V^{\rmnum{3}}_{l_{123}}\right) -  V^{\rmnum{2}}_{l_{123}} + 
V^{\rmnum{4}}_{l_{123}} + 2\sin^2(2\bar{\alpha}) V^{\rmnum{5}}_{l_{123}}  \Big]\nn\\
&& + \frac{1}{6}\big[\mathcal{E}_{l_{\mathrm{sum}}} (-1)^{l_{\mathrm{sum}}/2} +  i\mathcal{O}_{l_{\mathrm{sum}}} (-1)^{(l_{\mathrm{sum}}-1)/2} \big] h^{-1}_{l_1l_2l_3} \sum_{L_i, i=1}^3 \sin\left( \beta_1\right)\sin\left( \beta_2\right)\sin\left( \beta_3\right)V^{\rmnum{6}}(L_{123}, l_{123})\nn\\
&& + (5\mathrm{~perms}),
\label{bispectra_eee}
\eea
where the angle parameters $\beta_i$ and the auxiliary functions $V_{l_{123}}$are defined as
\bea
\beta_1 &=& \frac{L_1+L_3+l_1}{2}\pi-2\bar{\alpha}, ~~ \beta_2 =\frac{L_2+L_3+l_2}{2}\pi-2\bar{\alpha}, ~~
\beta_3 = \frac{L_1+L_2+l_3}{2}\pi-2\bar{\alpha}~,\nn
\eea
\bea
V^0_{l_{123}} &=& \frac{k}{4} C_{l_2}^{EE}C_{l_3}^{E\alpha} , \nn\\
V^{\rmnum{1}}_{l_{123}} &=&   \frac{ k C_{l_3}^{E\alpha} }{(2l_1+1) }    \sum_{L_i, i=1}^2    I_{L_2L_1l_1}^{0-22}  I_{L_1L_2l_1}^{0-22}  C_{L_1}^{E\alpha}C_{L_2}^{{E\alpha}}, \nn\\ 
V^{\rmnum{2}}_{l_{123}}  &=&  \frac{ k C_{l_3}^{E\alpha} }{(2l_1+1) }  \sum_{L_i, i=1}^2    I_{L_2L_1l_1}^{0-22}  I_{L_1L_2l_1}^{0-22}  C_{L_1}^{E\alpha}C_{L_2}^{{E\alpha}}  (-1)^{L_1+L_2+l_1}, \nn\\
V^{\rmnum{3}}_{l_{123}}  &=&  \frac{ k C_{l_3}^{{E\alpha}} } {(2l_2+1) }      \sum_{L_i, i=1}^2   I_{L_1L_2l_2}^{0-22}  I^{0-22}_{L_1L_2l_2}  C_{L_1}^{\alpha\alpha} C_{L_2}^{EE}  (-1)^{L_1+L_2+l_2}, \nn\\
V^{\rmnum{4}}_{l_{123}} &=&  \frac{k C_{l_3}^{{E\alpha}} } {(2l_2+1) }    \sum_{L_i, i=1}^2   I_{L_1L_2l_2}^{0-22}  I^{0-22}_{L_1L_2l_2}  C_{L_1}^{\alpha\alpha} C_{L_2}^{EE} ,\nn\\
V^{\rmnum{5}}_{l_{123}} &=&   \frac{ k C_{l_2}^{EE}}{(2l_3+1) } \sum_{L_i, i=1}^2  I^{000}_{L_1L_2l_3} I_{L_1L_2l_3}^{0-22}    C_{L_1}^{\alpha\alpha}     C_{L_2}^{{E\alpha}} , \nn
\eea
\bea
V^{\rmnum{6}}(L_{123}, l_{123})  &=& 8e^{-6C^\alpha(0)}  \left\{ {\begin{array}{ccc}
	l_1 & l_2 & l_3  \\
	L_2 & L_1 & L_3  \\
	\end{array} } \right\}  \Big\{ f ( L_{123}, l_{123})  + f(L_{312}, l_{231}) +  f(L_{231}, l_{312})+ g ( L_{123}, l_{123}) \nn\\
&+&  (-1)^{l_{\mathrm{sum}}}  \left[ f(L_{213}, l_{213})  +   f(L_{132}, l_{321})   +   f(L_{321}, l_{132}) +g ( L_{321}, l_{132}) \right] \Big\}, 
\label{auxiliary_PPP}
\eea
here we denote $k=4 e^{-6C^\alpha(0)} I^{0-22}_{l_3l_1l_2}$. The function $f, g$ are defined as follows:
\bea
f ( L_{123}, l_{123})&=&  (-1)^{L_3+L_2+l_1+l_3} I_{L_1L_3l_1}^{0-22}  I_{L_2L_3l_2}^{0-22}I_{L_2L_1l_3}^{0-22} C_{L_3}^{EE}  C_{L_2}^{\alpha\alpha}C_{L_1}^{{E\alpha}}, \nn\\
g ( L_{123}, l_{123}) &=&  I_{L_1L_3l_1}^{0-22} I_{L_3L_2l_2}^{0-22}I_{L_2L_1l_3}^{0-22} (-1)^{ l_{\mathrm{sum}}}  C_{L_3}^{E\alpha}  C_{L_2}^{{E\alpha}}C_{L_1}^{{E\alpha}}, \nn\\
g ( L_{123}, l_{123}) &=& g ( L_{312}, l_{231}) = g ( L_{231}, l_{312}), \nn\\
g ( L_{213}, l_{213}) &=& g ( L_{321}, l_{132} ) = g ( L_{132}, l_{321})~.
\eea

Finally, we obtain the full expressions for the rest bispectra $EEB, EBB, BBB$:
\bea
\widetilde{b}^{EEB}_{l_1l_2l_3} &=&    \mathcal{E}_{l_{\mathrm{sum}}} \cos(2\bar{\alpha}) \Bigg\{2 \bigg[\cos^2(2\bar{\alpha}) V^0_{l_{321}}   -\sin^2(2\bar{\alpha})  \left(  V^0_{l_{123}}+  V^0_{l_{132}} \right)    \bigg]
+  V^{\rmnum{2}}_{l_{132}} - V^{\rmnum{2}}_{l_{321}} 
+  V^{\rmnum{4}}_{l_{321}} - V^{\rmnum{4}}_{l_{132}} 
  \nn\\
&&+\bigg[2 \sin^2(2\bar{\alpha})  V^{\rmnum{1}}_{l_{123}}- \cos(4\bar{\alpha})   \left(  V^{\rmnum{1}}_{l_{321}} +    V^{\rmnum{1}}_{l_{132}}\right) \bigg] + \bigg[ 2\sin^2(2\bar{\alpha}) V^{\rmnum{3}}_{l_{123}}- \cos(4\bar{\alpha})    \left( V^{\rmnum{3}}_{l_{321}} - V^{\rmnum{3}}_{l_{132}}\right)  \bigg]  \nn\\
&& + 2\bigg[- \cos^2(2\bar{\alpha}) V^{\rmnum{5}}_{l_{123}}+\sin^2(2\bar{\alpha})   \left(\ V^{\rmnum{5}}_{l_{321}}  +  V^{\rmnum{5}}_{l_{132}}   \right)  \bigg]   \Bigg\}h^{-1}_{l_1l_2l_3}  
+i \mathcal{O}_{l_{\mathrm{sum}}}\sin(2\bar{\alpha})   \Bigg\{
- V^{\rmnum{2}}_{l_{123}}+V^{\rmnum{4}}_{l_{123}} \nn\\
&&+  \bigg[ -\cos(4\bar{\alpha}) 
V^{\rmnum{3}}_{l_{123}} +2 \cos^2(2\bar{\alpha})\left( V^{\rmnum{3}}_{l_{321}} + V^{\rmnum{3}}_{l_{132}} \right)  \bigg] + 2\bigg[  -   \sin^2(2\bar{\alpha}) V^{\rmnum{5}}_{l_{132}} +\cos^2 (2\bar{\alpha})\left(    V^{\rmnum{5}}_{l_{321}} - V^{\rmnum{5}}_{l_{123}} \right)     \bigg] \nn\\
&&+ 2 \cos^2(2\bar{\alpha}) \bigg[   V^0_{l_{123}}-  V^0_{l_{321}}-   V^0_{l_{132}}    \bigg] + \bigg[  \cos(4\bar{\alpha})V^{\rmnum{1}}_{l_{123}}- 2\cos^2(2\bar{\alpha}) \left(   V^{\rmnum{1}}_{l_{321}}+   V^{\rmnum{1}}_{l_{132}}\right)  \bigg] \Bigg\} h^{-1}_{l_1l_2l_3}   \nn\\
&& +  \frac{1}{2  }\big[\mathcal{E}_{l_{\mathrm{sum}}} (-1)^{l_{\mathrm{sum}}/2} +  i\mathcal{O}_{l_{\mathrm{sum}}} (-1)^{(l_{\mathrm{sum}}-1)/2} \big]  h^{-1}_{l_1l_2l_3}\sum_{L_2L_3L_1}  \sin\left( \beta_1\right)\sin\left( \beta_2\right)\cos\left( \beta_3\right)   V^{\rmnum{6}}(L_{123}, l_{123})\nn\\
&& + (l_1 \leftrightarrow l_2)~,  \label{bispectra_eeb}
\eea
\bea
\widetilde{b}^{EBB}_{l_1l_2l_3} 
&=&  \mathcal{E}_{l_{\mathrm{sum}}} \sin(2\bar{\alpha})\Bigg\{2\bigg[-  \sin^2(2\bar{\alpha})V^0_{l_{123}}+\cos^2(2\bar{\alpha}) \left(  V^0_{l_{213}} +  V^0_{l_{321}}\right)\bigg]+ 
+ V^{\rmnum{2}}_{l_{123}} - V^{\rmnum{2}}_{l_{213}}
+ V^{\rmnum{4}}_{l_{213}} - V^{\rmnum{4}}_{l_{123}}\nn\\
&& \bigg[ 2\cos^2(2\bar{\alpha}) V^{\rmnum{1}}_{l_{321}} - \cos(4\bar{\alpha}) \left( V^{\rmnum{1}}_{l_{123}}+ V^{\rmnum{1}}_{l_{213}}\right) \bigg] -\bigg[2\cos^2(2\bar{\alpha})V^{\rmnum{3}}_{l_{321}} + \cos(4\bar{\alpha}) \left( V^{\rmnum{3}}_{l_{123}}+ V^{\rmnum{3}}_{l_{213}}\right) \bigg] \nn\\
&&  + 2\bigg[\sin^2(2\bar{\alpha})  V^{\rmnum{5}}_{l_{321}}   -  \cos^2(2\bar{\alpha})  \left( V^{\rmnum{5}}_{l_{123}} +  V^{\rmnum{5}}_{l_{213}} \right) \bigg] \Bigg\} h^{-1}_{l_1l_2l_3}
+i \mathcal{O}_{l_{\mathrm{sum}}} \cos(2\bar{\alpha}) \Bigg\{-V^{\rmnum{2}}_{l_{321}} - V^{\rmnum{4}}_{l_{321}}  \nn\\
&& -\bigg[\cos(4\bar{\alpha})  V^{\rmnum{3}}_{l_{321}} + 2\sin^2(2\bar{\alpha})\left( V^{\rmnum{3}}_{l_{123}}- V^{\rmnum{3}}_{l_{213}}\right) \bigg] -2\bigg[\cos^2(2\bar{\alpha}) V^{\rmnum{5}}_{l_{213}} +  \sin^2(2\bar{\alpha})  \left(V^{\rmnum{5}}_{l_{123}} -  V^{\rmnum{5}}_{l_{321}}\right) \bigg]  \nn\\
 &&2 \sin^2(2\bar{\alpha})\bigg[  V^0_{l_{123}}-  V^0_{l_{213}}- V^0_{l_{321}} \bigg]
 -\bigg[\cos(4\bar{\alpha})V^{\rmnum{1}}_{l_{321}} -2\sin^2(2\bar{\alpha}) \left( V^{\rmnum{1}}_{l_{123}}- V^{\rmnum{1}}_{l_{213}}\right) \bigg] \Bigg\} h^{-1}_{l_1l_2l_3}  \nn\\
&& + \frac{1}{2  } \big[\mathcal{E}_{l_{\mathrm{sum}}} (-1)^{l_{\mathrm{sum}}/2} +  i\mathcal{O}_{l_{\mathrm{sum}}} (-1)^{(l_{\mathrm{sum}}-1)/2} \big]  h^{-1}_{l_1l_2l_3}\sum_{L_2L_3L_1}   \sin\left( \beta_1\right)\cos\left( \beta_2\right)\cos\left( \beta_3\right)   V^{\rmnum{6}}( L_{123}, l_{123})\nn\\
&&+  (l_2 \leftrightarrow l_3)~, \label{bispectra_ebb}
\eea
\bea
\widetilde{b}^{BBB}_{l_1l_2l_3} &=&- 2 \mathcal{E}_{l_{\mathrm{sum}}}h^{-1}_{l_1l_2l_3} \cos(2\bar{\alpha}) \sin^2(2\bar{\alpha})  \Big(-V^{0}_{l_{123}}+ V^{\rmnum{1}}_{l_{123}}+ V^{\rmnum{3}}_{l_{123}}+V^{\rmnum{5}}_{l_{123}}\Big)\nn\\
&& + i \mathcal{O}_{l_{\mathrm{sum}}} h^{-1}_{l_1l_2l_3}\sin(2\bar{\alpha}) \Big[ 2\sin^2(2\bar{\alpha})V^0_{l_{123}} + \cos(4\bar{\alpha}) \left( V^{\rmnum{3}}_{l_{123}}    - V^{\rmnum{1}}_{l_{123}}\right) - V^{\rmnum{2}}_{l_{123}}+ V^{\rmnum{4}}_{l_{123}} + 2\cos^2(2\bar{\alpha})V^{\rmnum{5}}_{l_{123}}  \Big]\nn\\ 
&&  +\frac{1}{6} \big[\mathcal{E}_{l_{\mathrm{sum}}} (-1)^{l_{\mathrm{sum}}/2} +  i\mathcal{O}_{l_{\mathrm{sum}}} (-1)^{(l_{\mathrm{sum}}-1)/2} \big] 
h^{-1}_{l_1l_2l_3}\sum_{L_i, i=1}^3      \cos\left( \beta_1\right)\cos\left( \beta_2\right)\cos\left( \beta_3\right) V^{\rmnum{6}}(L_{123}, l_{123})\nn\\
&& + (5\mathrm{~perms})~.\label{bispectra_bbb}~. 
\eea
we see again that to have non-vanishing rotated $PPP$ bispectra, the $E\alpha$ correlation is necessary and both parity-odd and parity-even terms are present in these expressions. 

To get a little bit more intuition of how big these reduced bispectra are, we try to find the approximate relationship between them and the small variables $\bar{\alpha}$.
Notice the auxiliary function $V^{0}_{l_{123}}$ is proportional to $C_l^{E\alpha}$ so that it is of the first order of parameter $\epsilon$, and the rest of auxiliary functions from $V_{l_{123}}^I$ to $V_{l_{123}}^{VI}$ are at least two orders higher so that they can be neglected. 
After preserving the leading order, we get a rough approximations as follows (explicit expressions can be found eq.~(\ref{approximate_ppp})),
\bea 
\widetilde{b}^{EEE}(\mathrm{even}) \sim   \bar{\alpha}    C_{l}^{E\alpha} &, ~~ & \widetilde{b}^{EEE}(\mathrm{odd}) \sim    C_{l}^{E\alpha}, \nn\\
\widetilde{b}^{EEB}(\mathrm{even}) \sim   C_l^{E\alpha} &, ~~ & \widetilde{b}^{EEB}(\mathrm{odd}) \sim   \bar{\alpha}  C_l^{E\alpha}, \nn\\
\widetilde{b}^{EBB}(\mathrm{even}) \sim  \bar{\alpha}C_l^{E\alpha}&, ~~ & \widetilde{b}^{EBB}(\mathrm{odd}) \sim (  \bar{\alpha}^2C_l^{E\alpha}, C_{L_1}^{E\alpha}C_{L_2}^{\alpha\alpha}, C_{L_1}^{E\alpha}C_{L_2}^{E\alpha}C_{L_3}^{E\alpha}), \nn\\
\widetilde{b}^{BBB}(\mathrm{even}) &\sim & ( \bar{\alpha}^2C_l^{E\alpha},   C_{L_1}^{E\alpha}C_{L_2}^{\alpha\alpha}, C_{L_1}^{E\alpha}C_{L_2}^{E\alpha}C_{L_3}^{E\alpha}),  \nn\\
\widetilde{b}^{BBB}(\mathrm{odd}) &\sim  &  ( \bar{\alpha}^3C_l^{E\alpha},   \bar{\alpha}C_{L_1}^{E\alpha}C_{L_2}^{\alpha\alpha}, \bar{\alpha}C_{L_1}^{E\alpha}C_{L_2}^{E\alpha}C_{L_3}^{E\alpha}).
\label{PPP_1st}
\eea

\begin{figure}[t]
	\centering
	\includegraphics[trim=4.8cm 0 0 0, width=1.08\textwidth]{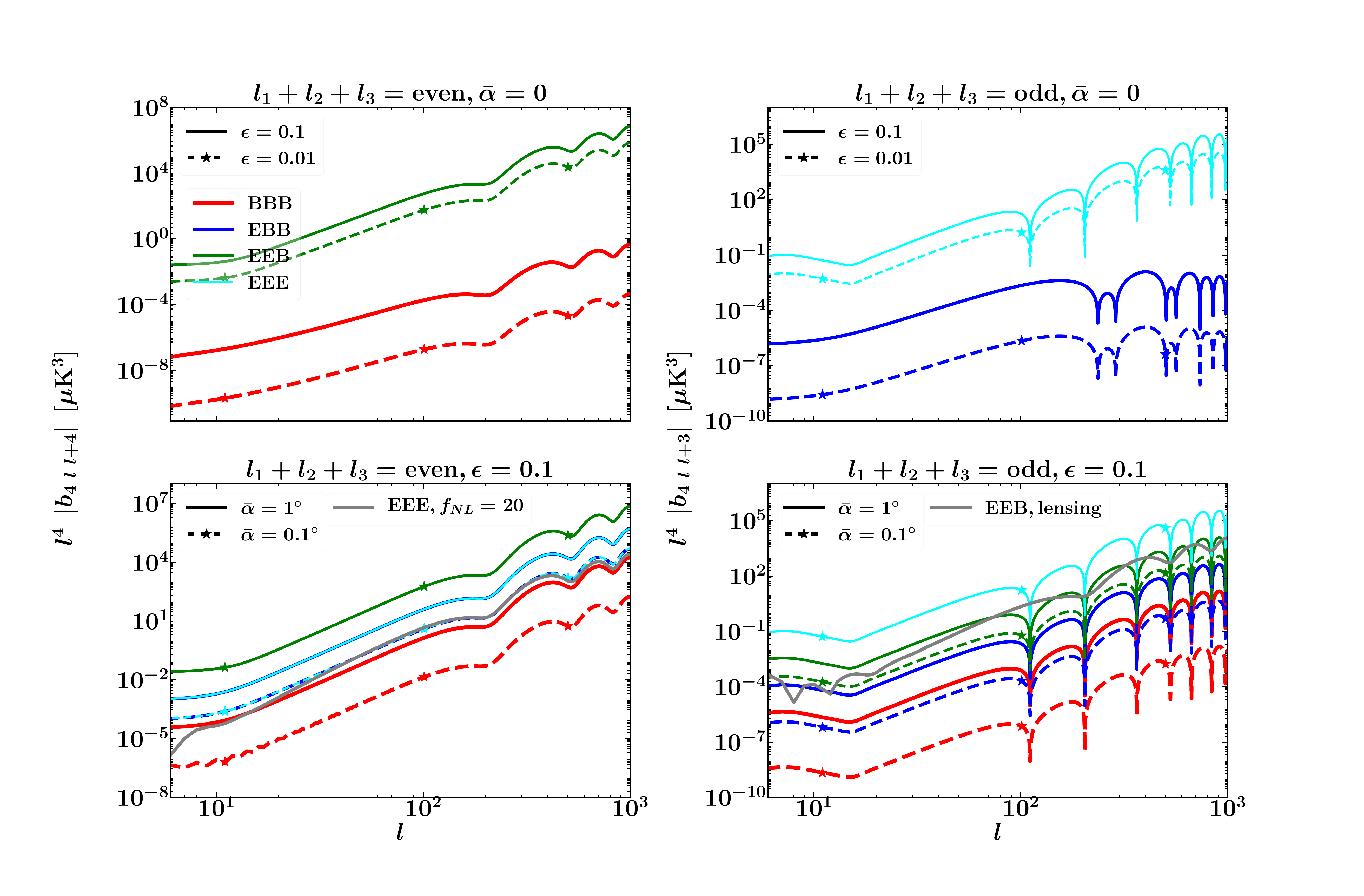}
	\caption{ Absolute values of rotated reduced bispectra  $b^{PPP}_{4l_2l_3}$ as functions of $l_2$ for various parameter and parities.  Configuration of multipole and parameter settings are as same as that in figure~\ref{bis_TTP}.  The primordial bispectra $  EEE  $ with  $f_{NL}=20$, and weak lensing induced $EEB$ are plotted for comparison.}
	\label{bis_PPP}	
\end{figure}

Analogue to the previous two cases, we plot the $PPP$ type bispectra in figure~\ref{bis_PPP}.
According to eq.~(\ref{PPP_1st}), odd-$EBB$ and even-$BBB$ are expected to be two orders smaller than  bispectra odd-$EEE$ and even-$EEB$. This is conformed in the figure. While for parity conserved parts, 
even-$EEE$ and even-$EBB$ have the same origin even-$EEB$, so that they are close to each other as the figure shows.  However, odd-$EEB$ is two orders larger than odd-$BBB$ although they are both rotated from odd-$EEE$ but the latter bispectrum has to rotate twice more.
As for the capacity of scientific interpretation, odd-$EEE$ and even-$EEB$ have potential to be powerful in probing anisotropic rotation angle, measuring $\epsilon$ and $C_l^{E\alpha}$, since both two bispectra are proportional to $C_l^{E\alpha}$.

\section{Conclusion and discussion}
\label{conclusion}

The Chern-Simons term, through which the cosmic Axion-like field couples to the electromagnetic field, has the effect to rotate CMB polarization directions and to break the CPT symmetry. There were lots of studies on this phenomenon focusing on the changes of the CMB power spectra by both the isotropic and anisotropic polarization rotation. However, almost all of these studies ignored the correlations between the (anisotropic) rotation angle $\alpha$ and the CMB temperature and (unrotated) $E$ polarization fields. These correlations could be generated in the Axion-like models with nonzero potential under the adiabatic initial condition. In this paper, we revisited the effects on CMB by the polarization rotation, taking into account the $T\alpha$ and $E\alpha$ correlations. We investigated the rotated CMB power spectra and bispectra. We found that $C_l^{T\alpha}$ has no contribution to the power spectra, but $C_l^{E\alpha}$ does, it brought a further modification to the $EE$, $BB$ and $EB$ power spectra relative to the previous results in ref. \cite{Li:2013vga}. This modification is one order of magnitude smaller than that by $C_l^{\alpha\alpha}$ at small multipoles about $l \le 200$.  When $l \ge 200$, the effect of $ C_l^{E\alpha}$  on power spectra can be neglected. 

The CMB bispectra vanish if the temperature and polarization fields are Gaussian. We found that this happens even for the rotated polarization fields if there are no $T\alpha$ and/or $E\alpha$ correlations.  By considering non-zero $C_l^{T\alpha}$ and $C_l^{E\alpha}$, we calculated  the rotated CMB bispectra analytically.  Besides their dependence on $C_l^{T\alpha}$ and $C_l^{E\alpha}$, these bispectra have the property to break parity symmetry. In their expressions, the isotropic rotation angle brings parity mixing and the anisotropic one induces parity odd bispectrum components. Then the measurement of parity odd bispectra is possible to be used to search CPT violation signals. Among these rotated bispectra, 
the produced $TTE$ and $TTB$ are proportional to $C_l^{T\alpha}$ thus can be used to make constraints on $T\alpha$ correlation. Similarly, the leading orders of produced $EEE$ and $EEB$ are proportional to $C_l^{E\alpha}$ and can also be used to constrain $E\alpha$ correlation. Since in the Axion-like model $C_l^{T\alpha}$ and $C_l^{E\alpha}$ are order one of Chern-Simons coupling parameter $\epsilon$ while  $C_l^{\alpha\alpha}$ is order two, rotated bispectra may provide a new efficient way to constrain  $\epsilon$. To date, non-Gaussianity on CMB temperature and $E$ polarization have been well measured by Planck \cite{Akrami:2019izv}. The constraints results on the nonlinear parameters of primordial tensor non-Gaussianity are $f_{NL}^{\mathrm{ten}}  = (1 \pm 18) \times 10^2$ from the the parity odd bispectra of $T+E$ map and $f_{NL}^{\mathrm{ten}} = (-570 \pm 720)\times 10^2$ from $E$ map only. 
The method estimating  $f_{NL}^{\mathrm{ten}}$ is also appropriate for parameter $\epsilon$ since the parity odd $TTE$ and the leading orders of parity odd $TEE, EEE$ are proportional to $\epsilon$. Furthermore, these parity odd bispectra can be used to reconstruct the $T\alpha$ and  $E\alpha$ correlations. To do so,  we should focus on the bispectra with special multipole configurations. For instance, fixing $l_1$ then $\widetilde{b}^{TTE}_{l_1l_2l_3} \sim C_{l_2}^{T\alpha}$ and by setting $l_1 = \mathrm{const}, l_2=l_3$ we have the leading order of $\widetilde{b}^{EEE}_{l_1l_2l_3} \sim C_{l_2}^{E\alpha}$. These work are based on tricks of estimating the bispectra from CMB map and we leave them in future.  Beyond the current temperature and $E$ polarization measurements, future observations of $B$ mode polarization from the experiments such as AliCPT \cite{Li:2017drr}, LiteBird \cite{Matsumura:2013aja}, PICO \cite{Hanany:2019lle} can enhance the detectability of the Chern-Simons coupling theory by utilization of $TTB, EEB$ bispectra data and so on.

\appendix

\section {Gaussian integration by parts}
\label{statistical_formulae}

Single scalar slow roll inflation theory predicts CMB temperature and polarization fields on the sky sphere are Gaussian distributed, meanwhile the statistics of CMB can be fully described by two point functions~\cite{Weinberg:2008zzc}.  But if the random field is not Gaussian distributed, there is no general way to calculate the $n$-point correlation functions. However as long as the $n$-point correlations are functions of multivariate Gaussian variables,  we can use the Gaussian integration by parts formula  ~\cite{vershynin_2018} to expand them into the products of two point functions.

For the zero-mean multivariate Gaussian random vector $x_1, ..., x_n$ ,  the formula is stated as
\bea
\llangle x_1 f(x_1, ...,x_n)\rrangle = \sum_i \llangle x_1 x_i\rrangle \llangle \partial_{x_i} f(x_1, ..., x_n) \rrangle
\label{gaussian_int_parts}  
\eea
where $f$ is the continuous functions of multiple gaussian variables.  By choosing different form of $f$, we can draw a lot of useful statistical formulae.

When $f$ is exponent function of gaussian variable $ e^{mx_2}$, $m$ is a constant, we get 
\bea
\llangle x_1e^{mx_2}\rrangle = \llangle x_1x_2\rrangle \llangle me^{mx_2}\rrangle  = m \langle x_1x_2\rangle  e^{m^2 \llangle x_2^2\rrangle/2}.
\label{n_exp}
\eea
here we used the expectation of log-normal distribution $ \llangle e^{mx} \rrangle = e^{m^2 \llangle x^2\rrangle/2 }$ \cite{janson_1997}. eq.~(\ref{n_exp}) is prepared for the following derivations.

In section \ref{section_power_spectrum}, the two point function of rotated polarization fields is actually the product of two unrotated CMB fields and one exponent function of rotation angle $\delta\alpha$. To solve that expectation, let $f = x_2e^{mx_3}$, we get corresponding expansion formula
\begin{eqnarray}
\llangle  x_1x_2e^{mx_3} \rrangle 
&=& e^{m^2 \llangle x_3^2\rrangle /2} \left[   \langle x_1x_2\rangle+ m^2\langle x_1x_3\rangle \langle x_2x_3\rangle\right],
\label{expectation2}
\end{eqnarray}

Similarly in section \ref{section_bispectrum}, rotated bispectra  contain three unrotated CMB fields and exponent function of the rotation angle $\delta\alpha$. Set $f= x_2x_3e^{mx_4}$,  we get 
\bea
\llangle x_1 x_2 x_3 e^{m x_4}\rrangle = m e^{m^2\llangle x_4^2\rrangle/2} \Big[  \langle x_1 x_2 x_3  x_4\rangle   + m^2 \langle x_4x_1\rangle\langle x_4x_2\rangle\langle x_4x_3\rangle\Big].
\label{expectation3}
\eea

We can also derive the expansion formula for the rotated trispectrum  by  setting $f = x_2x_3x_4e^{mx_5}$, then
\bea
\llangle x_1 x_2 x_3 x_4 e^{m x_5}\rrangle &=& e^{m^2 \llangle x_5^2\rrangle /2} \Big[ \llangle x_1 x_2 x_3 x_4 \rrangle + m^2 \llangle x_1 x_5 \rrangle\llangle  x_2 x_3 x_4 x_5 \rrangle  \nn\\
&+&  m^2  \llangle x_1 x_2 \rrangle \llangle x_3 x_5 \rrangle \llangle x_4 x_5 \rrangle + m^2  \llangle x_1 x_3 \rrangle \llangle x_2 x_5 \rrangle \llangle x_4 x_5 \rrangle  \nn\\ 
&+&   m^2  \llangle x_1 x_4 \rrangle \llangle x_2 x_5 \rrangle \llangle x_3 x_5 \rrangle + m^4\llangle x_1 x_5 \rrangle\llangle x_2 x_5 \rrangle  \llangle x_3 x_5 \rrangle \llangle x_4 x_5  \rrangle \Big].
\label{expectation4}
\eea

\section{Wigner D matrix,  spin spherical harmonics}
\label{spin-harmonic_property}

The simplification of formulae of rotated power spectra and bispectra  used the properties of both Wigner d matrix and spin weighted spherical harmonics.  Here we review these properties based on refs.~\cite{Goldberg:1966uu, Varshalovich:1988ye}.

For Wigner d matrix, it is related with Wigner D matrix as
\bea
D^l_{mm'}(\alpha, \beta, \gamma)  = e^{-im\alpha} d^l_{mm'}(\beta)e^{-im'\gamma}, ~~~ 0 \le \alpha \le 2\pi,~~ 0 \le \beta \le \pi, ~~0 \le \gamma \le 2\pi.
\eea
it has symmetries on the indices
\bea
d^l_{mm'}(\beta) &=& (-1)^{m-m'}d^l_{-m-m'}(\beta)= (-1)^{m-m'}d^l_{m'm}(\beta)=d^l_{-m'-m}(\beta), 
\eea
the Wigner 3j(Clebsch-Gorden) expansion for the product of two Wigner d matrix is 
\bea
d_{m'_1,m_1}^{l_1}( \beta )d_{m'_2,m_2}^{l_2}(\beta) &=& \sum_{l_3,m_3, m'_3}  (-1)^{m_3+m'_3}(2l_3+1) \nn\\
&\times&\left( {\begin{array}{ccc}
		l_1 &l_2 & l_3 \\
		m'_1 & m'_2 & -m'_3\\
\end{array} } \right)\left( {\begin{array}{ccc}
		l_1 &l_2 & l_3 \\
		m_1 & m_2 & -m_3\\
\end{array} } \right)  d_{m'_3, m_3}^{l_3}(\beta),
\label{small_d_expansion}
\eea

For spin weighted spherical harmonic functions on the sphere , it is  defined as, 
\bea
{_{s}}Y_{lm}(\theta, \phi) &=& \sqrt{\frac{2l+1}{4\pi}}D^l_{-sm}(\phi, \theta, \gamma) e^{is\gamma}, ~~~l \ge |s|.  
\eea
here $s$ is the spin number.

We can also expand product of two spin weighted spherical harmonics into the series of Wigner 3j symbols, 
\bea
{}_{s_1}{Y}{_{l_1 m_1}}(\hat{  n}) {}_{s_2}{Y}{_{l_2 m_2}}(\hat{  n})  &=& \sum_{l_3m_3s_3}  I_{l_1l_2l_3}^{-s_1,-s_2,-s_3} \left( {\begin{array}{ccc}
		l_1 &l_2 & l_3 \\
		m_1 & m_2 & m_3\\
\end{array} } \right) {}_{s_3}{Y} {^*_{l_3 m_3}}(\hat{  n}), 
\label{spin_harmonic_expansion}
\eea
where $I_{l_1l_2l_3}^{s_1s_2s_3}$ is Wigner 3j symbols defined in eq.~(\ref{wigner_i}).  

We have used the weak version of the addition theorem of spin weighted spherical harmonics, in order to eliminate the effects of $T\alpha$ correlations on power spectra and effects of $E\alpha$ correlation on $TTP$ bispectra,
\bea
\sum_{m=-l}^l {_{s_1}}Y_{lm}^*(\theta, \phi) {_{s_2}}Y_{lm}(\theta, \phi) &=&  \frac{2l+1}{4\pi}  \delta_{ss'} ,
\label{spin_addition_theorem}
\eea
this formula can be deduced from the unitary condition for Wigner D-matrix 
\bea
\sum_{m} D^{l*}_{sm}(\alpha, \beta, \gamma) D^{l}_{s'm}(\alpha, \beta, \gamma)  = \delta_{ss'},
\eea

A lot of calculations in this work are attributed to the gaunt function $\mathcal{G}_{l_1l_2l_3}^{m_1m_2m_3}$, i.e., the integration of  product of triple spherical harmonics over 2-d sphere, 
\bea
\int d\Omega\;   {}_{s_1}Y_{l_1m_1}(\bm n)  {}_{s_2}Y_{l_2m_2}(\bm n)  {}_{s_3}Y_{l_3m_3}(\bm n)  
&=&  I^{-s_1-s_2-s_3}_{l_1l_2l_3}  \left( {\begin{array}{ccc}
		l_1 & l_2 & l_3 \\
		m_1 & m_2 & m_3\\
\end{array} } \right).
\label{gaunt_formula}
\eea

\section{Wigner symbols}
\label{wigner_property}

From eq.~(\ref{gaunt_formula}),  integrations of product of spin weighted harmonic can be simplified into expressions of Wigner 3j symbols. We then can make further simplifications with the properties of Wigner 3j/6j symbols.  The following formulae are based on ref. \cite{Varshalovich:1988ye}, they were used in derivation of the $PPP$ bispectra.

Wigner 3j symbol is related with the Clebsch-Gorden coefficient by
\bea
\llangle  l_1m_1l_2m_2|l_3m_3\rrangle  = (-1)^{l_1-l_2+m_3}\sqrt{2l_3+1} \left( {\begin{array}{ccc}
		l_1 &l_2 & l_3 \\
		m_1 & m_2 & -m_3\\
\end{array} } \right)  
\eea
The symbol satisfies the selection rule and triangle condition
\bea
|m_i|\le l_i,~~ i=1,2,3;~~ m_1+m_2= m_3; ~~|l_2-l_3|\le l_1 \le |l_2+l_3|.
\eea

When two angular momenta are equal $l_1=l_2$, Wigner 3j symbol simplifies
\bea
\left( {\begin{array}{ccc}
		l & l & 0 \\
		m & -m & 0\\
\end{array} } \right) &=& \frac{(-1)^{l-m}}{\sqrt{2l+1}}, 
\eea

Wigner 3j symbols satisfy the orthogonality relations
\bea
\sum_{l_3m_3}(2l+1)  \left( {\begin{array}{ccc}
		l_1 &l_2 & l_3\\
		m_1 & m_2 & m_3\\
\end{array} } \right)  \left( {\begin{array}{ccc}
		l_1 &l_2 & l_3\\
		m_1' & m_2' & m_3\\
\end{array} } \right) &=& \delta_{m_1 m_1'} \delta_{m_2 m_2'}\nn\\
\sum_{m_1 m_2}  \left( {\begin{array}{ccc}
		l_1 &l_2 & l_3\\
		m_1 & m_2 & m_3\\
\end{array} } \right)  \left( {\begin{array}{ccc}
		l_1 &l_2 & l'_3 \\
		m_1 & m_2 & m'_3\\
\end{array} } \right) &=& \frac{1}{2l_3+1}\delta_{l_3 l_3'} \delta_{m_3m'_3}
\eea

Wigner 6j symbol is related with different coupling schemes of three angular momenta, consider
\bea
&&\bm {l_1}+ \bm{l_2} = \bm {l_{3}},  ~~~\bm {l_{3}}+ \bm {l_4} = \bm {l_5},\nn\\
&& \bm {l_2}+ \bm{l_4} = \bm {l_{6}},  ~~~\bm {l_1}  + \bm {l_{6}}= \bm {l_5}.
\label{coupling_momentas}
\eea
then Wigner 6j symbol is defined as
\bea
\llangle l_1l_2(l_{3})l_4l_5m_5|l_1,l_2l_4(l_{6})l_5m_5\rrangle  = (-1)^{l_1+l_2+l_4+l_5}\sqrt{(2l_{3}+1)(2l_{6}+1)} \left\{ {\begin{array}{ccc}
		l_1 &l_2 & l_{3} \\
		l_4 & l_5 & l_{6}\\
\end{array} } \right\}
\eea
eq. (\ref{coupling_momentas}) contains the triangle conditions for 6j symbol.

The 6j symbol can be expressed by the 3j symbols,  
\bea
\left( {\begin{array}{ccc}
		l_1 &l_2 & l_3 \\
		m_1 & m_2 & m_3\\
\end{array} } \right)&&\left\{ {\begin{array}{ccc}
		l_1 &l_2 & l_3 \\
		l_4 & l_5 & l_6\\
\end{array} } \right\} =\sum_{m_4m_5m_6}(-1)^{l_4-m_4+l_5-m_5+l_6-m_6} \nn\\
\times&&\left( {\begin{array}{ccc}
		l_5 &l_1 & l_6 \\
		m_5 & -m_1 & -m_6\\
\end{array} } \right)\left( {\begin{array}{ccc}
		l_6 &l_2 & l_4 \\
		m_6 & -m_2 & -m_4\\
\end{array} } \right)\left( {\begin{array}{ccc}
		l_4 &l_3 & l_5 \\
		m_4 & -m_3 & -m_5\\
\end{array} } \right)
\eea
applying the orthogonality formula of 3j symbols to this equation, we obtain the summation formula, 
\begin{eqnarray}
&&\sum_{l_6}(2l_6+1)(-1)^{l_6-m_6}\left( {\begin{array}{ccc}
	l_1 & l_5 & l_6 \\
	m_1 & m_5 & m_6\\
	\end{array} }\right) \left( {\begin{array}{ccc}
	l_6 & l_4 & l_2   \\
	-m_6 & m_4 & m_2 \\
	\end{array} } \right)\left\{ {\begin{array}{ccc}
	l_1 & l_2 & l_3 \\
	l_4 & l_5 & l_6\\
	\end{array} }\right\} \nn\\
&=& \sum_{m_3} (-1)^{l_3-m_3}\left( {\begin{array}{ccc}
	l_1 & l_2 & l_3 \\
	m_1 & m_2 & m_3\\
	\end{array} }\right)\left( {\begin{array}{ccc}
	l_3 & l_4 & l_5    \\
	-m_3 & m_4 & m_5\\
	\end{array} }\right)
\end{eqnarray}

\section{Simplification of integration over two directions}
\label{diagonal_proof}

eq.~(\ref{spin_harmonic_expansion}) tells us the conjugate product of two same-spin weighted spherical harmonics can be transformed into ordinary spherical harmonics, hence the total integration in eqs.~(\ref{BB_power}, \ref{TPP_type_int}) can be simplified and attributed to the type of expression  
\bea
\int d\hat{\bm n_1} \int d\hat{\bm n_2} \cdot  {Y}_{l_1 m_1}^*(\hat{\bm  n_1}) {Y}{_{l_2 m_2}}(\hat{ \bm  n_2}) e^{kC^\alpha ({\bm n_1 \cdot \bm n_2})}, 
\label{wigner1_diagonal}
\eea
here $k$ is a constant number. Apparently, the two directions in the integrand are not separable and there seems no direct way to simplify the integration. Note the integrand is symmetric on the two directions,  we can prove that the above expression is diagonal with indices $l_1m_1, l_2m_2$, i.e., 
\bea
\int d\hat{\bm n_1} \int d\hat{\bm n_2} \cdot  {Y}_{l_1 m_1}^*(\hat{\bm  n_1}) {Y}{_{l_2 m_2}}(\hat{ \bm  n_2}) e^{kC^\alpha ({\bm n_1 \cdot \bm n_2})} \sim f(l_1)\delta_{l_1l_2}\delta_{m_1m_2} , 
\label{diagonal_integration}
\eea
where $f(l_1)$ is the sole function of $l_1$. This conclusion is useful because the integration doesn't depend on the indices $m_1,m_2$, so we can use the addition theorem of spherical harmonic to make average on $m$ and hence could get a more concise result.

The idea of the proof is to expand the exponent into series and analyze the results of arbitrary order. If eq. (\ref{diagonal_integration}) is valid for each order, then it is proved. Taylor expand the exponent part, apply eq. (\ref{alpha_correlation}) and write the lengendre function into product of spherical harmonics, we obtain 
\bea
e^{kC^\alpha ({\bm n_1 \cdot \bm n_2})} &=& 1 + k\sum_{LM}C_L^{\alpha\alpha}Y_{LM}(\bm n_1)Y^*_{LM}(\bm n_2) + ...\nn\\
&+& \frac{k^n}{n!}\sum_{L_iM_i,  i=1}^{ i=n}\prod_{i=1}^n C_{L_i}^{\alpha\alpha}Y_{L_iM_i}( \bm n_1)Y_{L_iM_i}^*(\bm n_2)+ ...
\eea

The integration corresponding to the zero-th and first order can be obtained analytically, 
\bea
\int d\hat{\bm n_1} \int d\hat{\bm n_2} \cdot  {Y}_{l_1 m_1}^*(\hat{\bm  n_1}) {Y}{_{l_2 m_2}}(\hat{ \bm  n_2}) e^{kC^\alpha ({\bm n_1 \cdot \bm n_2})} = \left\{\begin{array}{rcl}
	4\pi \delta_{l_1l_2}\delta_{m_1m_2} & & {\mathrm{zero-th~order}}\\
	kC_{l_1}^{\alpha\alpha}\delta_{l_1l_2}\delta_{m_1m_2} & & {\mathrm{first~order}}
\end{array}\right.~.
\eea
obviously results of first two order are consistent with  (\ref{diagonal_integration}).

For arbitrary order check,  because the $n$-th expansion contains product of $2n$ spherical harmonics, we first use eq. (\ref{spin_harmonic_expansion}) to transform two harmonic into one. For example,
\bea
&&e^{kC^\alpha(\bm n_1\cdot \bm n_2)}(n) \nn\\
&=& \frac{k^n}{n!}\sum_{L_iM_i, i=1}^{2}  C_{L_1}^{\alpha\alpha}  C_{L_2}^{\alpha\alpha} \overbrace{Y_{L_1M_1}( \bm n_1)Y_{L_2M_2}( \bm n_1)} \underbrace{ Y_{L_1M_1}^*(\bm n_2) Y_{L_2M_2}^*(\bm n_2)}  \sum_{L_iM_i, i=3}^{n}\prod_{i=3}^n C_{L_i}^{\alpha\alpha}Y_{L_iM_i}( \bm n_1)Y_{L_iM_i}^*(\bm n_2)\nn\\
&=&  \frac{k^n}{n!}\sum_{L_i, i=1}^{2}  C_{L_1}^{\alpha\alpha}C_{L_2}^{\alpha\alpha}    \sum_{L'_2M'_2}\sum_{L''_2M''_2} I^{000}_{L_1L_2L'_2} I^{000}_{L_1L_2L''_2}\sum_{M_i, i=1}^{2}\left( {\begin{array}{ccc}
		L_1 &L_2 & L'_2 \nn\\
		M_1 & M_2 & M'_2\nn\\
\end{array} } \right) \left( {\begin{array}{ccc}
		L_1 &L_2 & L''_2 \nn\\
		-M_1 & -M_2 & M''_2\nn\\
\end{array} } \right) \nn\\
&& \times  (-1)^{M_1+M_2+M'_2}  Y_{L'_2-M'_2}(\bm n_1) Y_{L''_2M''_2}^*(\bm n_2)  \sum_{L_iM_i, i=3}^{n}\prod_{i=3}^n C_{L_i}^{\alpha\alpha}Y_{L_iM_i}( \bm n_1)Y_{L_iM_i}^*(\bm n_2),\nn\\
&=&  \frac{k^n}{n!}\sum_{L_i, i=1}^{2}  C_{L_1}^{\alpha\alpha}C_{L_2}^{\alpha\alpha}    \sum_{L'_2M'_2} f_{L_1L_2L'_2}    Y_{L'_2M'_2}(\bm n_1) Y_{L'_2M'_2}^*(\bm n_2)  \sum_{L_iM_i, i=3}^{n}\prod_{i=3}^n C_{L_i}^{\alpha\alpha}Y_{L_iM_i}( \bm n_1)Y_{L_iM_i}^*(\bm n_2),
\eea 
where $f_{L_1L_2L'_2}= (-1)^{L_1+L_2+L'_2}( I^{000}_{L_1L_2L'_2})^2  /(2L'_2+1) $. The last step used the orthogonality relation of Wigner 3j symbol.  Repeat this procedure until there left only two spherical harmonics,  we have
\begin{eqnarray}
e^{kC^\alpha(\bm n_1\cdot \bm n_2)}(n) 
&=& \frac{k^n}{n!}\sum_{L_i, i=1}^{n} \prod_{i=1}^n C_{L_i}^{\alpha\alpha} \sum_{L'_i, i=2}^n \prod_{i=2}^n f_{L'_{i-1}L_{i}L'_{i}} \sum_{M'_n} Y_{L'_nM'_n}(\bm n_1) Y_{L'_nM'_n}^*(\bm n_2),
\label{reduced_norder}
\end{eqnarray}
where $L'_1 = L_1$ and $f_{L'_{i-1}L_iL'_i}= (-1)^{L'_{i-1}+L_i+L'_i}( I^{000}_{L'_{i-1}L_iL'_i})^2  /(2L'_i+1) $.

Substitute the reduced $n$-th expansion eq. (\ref{reduced_norder}) into eq.~(\ref{wigner1_diagonal}), from the orthogonality of spherical harmonics one directly obtain, 
\begin{eqnarray}
&&\int d\hat{\bm n_1} \int d\hat{\bm n_2} \cdot  {Y}_{l_1 m_1}^*(\hat{  \bm n_1}) {Y}{_{l_2 m_2}}(\hat{ \bm  n_2}) e^{kC^\alpha (\bm n_1 \cdot \bm n_2)}  (n)  \nn\\ 
&=&  \frac{k^n}{n!}\sum_{L_i, i=1}^{n} \prod_{i=1}^n C_{L_i}^{\alpha\alpha}   \sum_{L'_i, i=2}^n   \prod_{i=2}^n f_{L'_{i-1}L_iL'_i} \delta_{L'_n l_1} \delta_{l_1 l_2}\delta_{m_1 m_2}.
\label{nth_expansion}
\end{eqnarray}
since $f_{L'_{i-1}L_iL'_i}$ doesn't depend on the indices $m$ so conclusion (\ref{diagonal_integration}) is valid for arbitrary order of the exponent. Hence it is proved.

Although we get the explicit result for arbitrary order expansion, the summation of all the orders' results is not easy to calculate. Nevertheless, consider the integration is identical for every $-l_1\le m_1 \le l_1$, make average over $m$ and use the addition theorem of spherical harmonics, we obtain
\bea
&&\int d\hat{\bm n_1} \int d\hat{\bm n_2} \cdot  {Y}^*_{l_1 m_1}(\bm{\hat{ n}}_1) {Y}_{l_2 m_2}(\bm{\hat{ n}}_2) e^{kC^\alpha (\bm n_1 \cdot \bm n_2)}    \nn\\
&=& \frac{1}{2l_1+1} \sum_{m_1= -l_1}^{l_1}\int d\hat{\bm n_1} \int d\hat{\bm n_2} \cdot  {Y}^*_{l_1 m_1}(\bm{\hat{ n}}_1) {Y}{_{l_1 m_1}}(\bm{\hat{ n}}_2) e^{kC^\alpha (\bm n_1 \cdot \bm n_2)} \delta_{l_1 l_2}\delta_{m_1 m_2}   \nn\\
&=& \frac{1}{4\pi}  \int d\hat{\bm n_1} \int d\hat{\bm n_2} \cdot P
_{l_1}(\bm n_1\cdot \bm n_2)e^{kC^\alpha (\bm n_1 \cdot \bm n_2)} \delta_{l_1 l_2}\delta_{m_1 m_2} \nn\\
&=& 2\pi \int_0^\pi \sin{\beta}d\beta~ P_{l_1}(\cos\beta) e^{kC^\alpha(\beta)}\delta_{l_1 l_2}\delta_{m_1 m_2} .
\label{exact_int_harmonics2} 
\eea
where $\cos\beta = \bm{n}_1\cdot \bm{n}_2$. The last step is calculated by choosing $\bm n_1 \parallel \bm z$.

Based on eq.~(\ref{exact_int_harmonics2}), we can solve  integration containing product of more than two spin weighted spherical harmonics functions
\bea
&&\sum_{q}\int d\hat{\bm n_1} \int d\hat{\bm n_2} \cdot {}_{s_1}{Y}{^*_{pq}}(\bm{\hat{ n}}_1) {}_{s_2}{Y}{_{pq}}(\bm{\hat{ n}}_2) {}_{s_1}{Y}{_{l_1 m_1}}(\bm{\hat{ n}}_1) {}_{s_2}{Y}^*{_{l_2 m_2}}(\bm{\hat{ n}}_2) e^{C^\alpha (\bm n_1 \cdot \bm n_2)}    \nn\\
&=&\frac{(2p+1)}{ 2} \int d\cos\beta \cdot d^{p}_{s_1s_2}(\beta)d^{l_1}_{s_1s_2}(\beta) e^{C^\alpha (\beta)}\delta_{l_1l_2}\delta_{m_1m_2}, 
\label{wigner2_diagonal}
\eea
\bea
&&\sum_{q_1q_2}\int d \bm{\hat{n}_1} \int d \bm{\hat{ n}}_2 e^{C^\alpha (\bm n_1 \cdot \bm n_2)} \times\nn\\
 &&\left[ {}_{s_1}{Y}{^*_{p_1q_1}}(\bm{\hat{ n}}_1) {}_{s_2}{Y}{_{p_1q_1}}(\bm{\hat{ n}}_2) {}_{s_3}{Y}{^*_{p_2q_2}}(\bm{\hat{ n}}_1) {}_{s_4}{Y}{_{p_2q_2}}(\bm{\hat{ n}}_2) {}_{s_1+s_3}{Y}{_{l_1 m_1}}(\bm{\hat{ n}}_1) {}_{s_2+s_4}{Y}^*{_{l_2 m_2}}(\bm{\hat{ n}}_2)\right]    \\
&& =\frac{(2p_1+1)(2p_2+1)}{8\pi} \int d\cos\beta   d^{p_1}_{s_1,s_2}(\beta)d^{p_2}_{s_3,s_4}(\beta) d^{l_1}_{s_1+s_3, s_2+s_4}(\beta) e^{C^\alpha (\beta)} \delta_{l_1l_2}\delta_{m_1m_2}. \nn
\label{wigner3_diagonal},
\eea
where we  used two wigner 3j expansion formulae for both Wigner d matrix and spin weighted spherical harmonics.

\section{Leading order approximations of reduced bispectra}
\label{complex_bispectra}

The polarization rotation angles are constrained to be vanishing small, hence it is more convenient to analyze the rotated bispectra from their leading order approximations rather than from the long and cumbersome expressions listed in section \ref{section_bispectrum}. Under the approximate condition eq.~(\ref{approximate_condition}) and use the tricks present on appendices ~(\ref{spin-harmonic_property}, \ref{wigner_property}), we obtain the leading order approximations for $TPP$ bispectra based on eqs.~(\ref{bispectra_tee}, \ref{TEB_bispectra}, \ref{TBB_bispectra})
\bea
\widetilde{b}^{TEE}_{l_1l_2l_3} & \approx &  
-    \mathcal{E}_{l_{\mathrm{sum}}} 4 \bar{\alpha} h^{-1}_{l_1l_2l_3} \Big[C_{l_1}^{TE} \left( I^{0-22}_{l_2l_3l_1}  C_{l_2}^{E\alpha} +I^{0-22}_{l_3l_2l_1}    C_{l_3}^{E\alpha} \right)  +   I_{l_1l_2l_3}^{0-22} C_{l_1}^{T\alpha}\left(  C_{l_3}^{EE}+  C_{l_2}^{EE}\right)\Big] \nn\\
&& + i\mathcal{O}_{l_{\mathrm{sum}}} 2h^{-1}_{l_1l_2l_3} \Big[ C_{l_1}^{TE}\left(  I^{0-22}_{l_2l_3l_1}    C_{l_2}^{E\alpha} - I^{0-22}_{l_3l_2l_1}    C_{l_3}^{E\alpha}\right) +   I_{l_1l_2l_3}^{0-22} C_{l_1}^{T\alpha}\left(  C_{l_3}^{EE} -  C_{l_2}^{EE}\right)  \Big], \nn\\
\label{approximate_tee}
\widetilde{b}^{TEB}_{l_1l_2l_3} & \approx &   \mathcal{E}_{l_{\mathrm{sum}}} 2h^{-1}_{l_1l_2l_3} \left(I^{0-22}_{l_2l_3l_1}   C_{l_1}^{TE}C_{l_2}^{E\alpha}     +   I_{l_1l_3l_2}^{0-22}C_{l_1}^{T\alpha}   C_{l_2}^{EE}  \right)\nn\\
&& +  i \mathcal{O}_{l_{\mathrm{sum}}}      4\bar{\alpha} h^{-1}_{l_1l_2l_3} \Big[  C_{l_1}^{TE}\left( I^{0-22}_{l_2l_3l_1}  C_{l_2}^{E\alpha} -I^{0-22}_{l_3l_2l_1}  C_{l_3}^{E\alpha}  \right)   + I_{l_1l_2l_3}^{0-22}         C_{l_1}^{T\alpha} \left(  C_{l_3}^{EE}   -     C_{l_2}^{EE}\right)  \Big] , \nonumber\\
\label{approximate_teb}
\widetilde{b}^{TBB}_{l_1l_2l_3}  & \approx &   \mathcal{E}_{l_{\mathrm{sum}}}  4\bar{\alpha} h^{-1}_{l_1l_2l_3}   \Big[C_{l_1}^{TE} \left( I^{0-22}_{l_2l_3l_1}  C_{l_2}^{E\alpha} +I^{0-22}_{l_3l_2l_1}    C_{l_3}^{E\alpha} \right)  +   I_{l_1l_2l_3}^{0-22} C_{l_1}^{T\alpha}\left(  C_{l_3}^{EE}+  C_{l_2}^{EE}\right)\Big] \nn\\
&&+     i\mathcal{O}_{l_{\mathrm{sum}}}    h^{-1}_{l_1l_2l_3} \Bigg\{ 2\bar{\alpha}^2 \Big[ C_{l_1}^{TE}\left( I^{0-22}_{l_2l_3l_1}         C_{l_2}^{E\alpha} - I^{0-22}_{l_3l_2l_1}         C_{l_1}^{TE}C_{l_3}^{E\alpha} \right)  +I_{l_1l_2l_3}^{0-22}  C_{l_1}^{T\alpha} \left(    C_{l_3}^{EE} - C_{l_2}^{EE}  \right)\Big]  \nn\\ 
&& +  4C_{l_1}^{T\alpha}I^{0-22}_{l_1l_2l_3}
\sum_{L_i, i=1}^2  C_{L_1}^{E\alpha }C_{L_2}^{E\alpha}  \bigg[ \frac{ I^{20-2}_{L_1L_2l_3} I^{0-22}_{L_1L_2l_3}\left( 1+(-1)^{L_1+L_2+l_3} \right)}{2l_3+1}\nn\\
&&  - \frac{ I^{20-2}_{L_1L_2l_2} I^{0-22}_{L_1L_2l_2}\left(  1+(-1)^{L_1+L_2+l_2} \right)}{2l_2+1}  \bigg] \Bigg\},	 
\label{approximate_tpp}
\eea

Similarly we obtain the leading order approximated results for $PPP$ bispectra based on eqs.~(\ref{bispectra_eee}, \ref{bispectra_eeb}, \ref{bispectra_ebb}, \ref{bispectra_bbb}), 
\bea
\widetilde{b}^{EEE}_{l_1l_2l_3} &\approx&  -    \mathcal{E}_{l_{\mathrm{sum}}}  4\bar{\alpha}  h^{-1}_{l_1l_2l_3} I^{0-22}_{l_3l_1l_2}    C_{l_2}^{EE}C_{l_3}^{E\alpha} +i \mathcal{O}_{l_{\mathrm{sum}}} 2 h^{-1}_{l_1l_2l_3} I^{0-22}_{l_3l_1l_2} C_{l_2}^{EE}C_{l_3}^{E\alpha}      + (5\mathrm{~perms}),\nn\\
\widetilde{b}^{EEB}_{l_1l_2l_3} &\approx&  i\mathcal{O}_{l_{\mathrm{sum}}} 4\bar{\alpha} h^{-1}_{l_1l_2l_3}\Big[I^{0-22}_{l_3l_1l_2} C_{l_2}^{EE}C_{l_3}^{E\alpha} - I^{0-22}_{l_1l_3l_2}  C_{l_2}^{EE}C_{l_1}^{E\alpha}  -  I^{0-22}_{l_2l_1l_3} C_{l_3}^{EE}C_{l_2}^{E\alpha}   \Big]\nn\\
&&+  \mathcal{E}_{l_{\mathrm{sum}}}  2h^{-1}_{l_1l_2l_3} I^{0-22}_{l_1l_3l_2}  C_{l_2}^{EE}C_{l_1}^{E\alpha}+ (l_1 \leftrightarrow l_2),\nn\\
\widetilde{b}^{EBB}_{l_1l_2l_3} &\approx&  \mathcal{E}_{l_{\mathrm{sum}}}  4 \bar{\alpha}h^{-1}_{l_1l_2l_3}\Big[I^{0-22}_{l_3l_2l_1}C_{l_1}^{EE}C_{l_3}^{E\alpha}+  I^{0-22}_{l_1l_3l_2} C_{l_2}^{EE}C_{l_1}^{E\alpha} \Big]\nn\\
&&+ i\mathcal{O}_{l_{\mathrm{sum}}}  8\bar{\alpha}^2 h^{-1}_{l_1l_2l_3}  \Big[ I^{0-22}_{l_3l_1l_2} C_{l_2}^{EE}C_{l_3}^{E\alpha}  - I^{0-22}_{l_3l_2l_1}C_{l_1}^{EE}C_{l_3}^{E\alpha} - I^{0-22}_{l_1l_3l_2} C_{l_2}^{EE}C_{l_1}^{E\alpha}  \Big] \nn\\
&& +  i \mathcal{O}_{l_{\mathrm{sum}}}  8h^{-1}_{l_1l_2l_3} \sum_{L_2L_3L_1}(-1)^{L_1+L_2+L_3} \left\{ {\begin{array}{ccc}
		l_1 & l_2 & l_3  \\
		L_2 & L_3 & L_1  \\
\end{array} } \right\} \Bigg[\nn\\
&& -I_{L_3L_1l_1}^{0-22}  I_{L_2L_1l_2}^{0-22}I_{L_2L_3l_3}^{0-22}\mathcal{O}_{L_3L_1l_1} \mathcal{E}_{L_2L_1l_2}    \mathcal{E}_{L_2L_3l_3} C_{L_1}^{EE} C_{L_2}^{\alpha\alpha} C_{L_3}^{{E\alpha}} \nn\\
&&- I_{L_2L_1l_2}^{0-22}  I_{L_3L_1l_1}^{0-22}I_{L_3L_2l_3}^{0-22}     \mathcal{E}_{L_2L_1l_2} \mathcal{O}_{L_3L_1l_1}    \mathcal{E}_{L_3L_2l_3} C_{L_1}^{EE}C_{L_3}^{\alpha\alpha}  C_{L_2}^{{E\alpha}} \nn\\
&& + I_{L_3L_2l_3}^{0-22}  I_{L_1L_2l_2}^{0-22}I_{L_1L_3l_1}^{0-22}      \mathcal{E}_{L_3L_2l_3} \mathcal{E}_{L_1L_2l_2}    \mathcal{O}_{L_1L_3l_1} C_{L_2}^{EE} C_{L_1}^{\alpha\alpha} C_{L_3}^{{E\alpha}}\nn\\
&&   - I_{L_3L_1l_1}^{0-22} I_{L_1L_2l_2}^{0-22}I_{L_2L_3l_3}^{0-22}   \mathcal{O}_{L_3L_1l_1} \mathcal{E}_{L_1L_2l_2}    \mathcal{E}_{L_2L_3l_3} C_{L_1}^{E\alpha}  C_{L_2}^{{E\alpha}}C_{L_3}^{{E\alpha}}  \Bigg]+ (l_2 \leftrightarrow l_3),
\nn
\eea
\bea
\widetilde{b}^{BBB}_{l_1l_2l_3} &\approx&  \mathcal{E}_{l_{\mathrm{sum}}}  8\bar{\alpha}^2 h^{-1}_{l_1l_2l_3} I^{0-22}_{l_3l_1l_2}   C_{l_2}^{EE}C_{l_3}^{E\alpha}   + i  \mathcal{O}_{l_{\mathrm{sum}}} 16\bar{\alpha}^3 h^{-1}_{l_1l_2l_3}  I^{0-22}_{l_3l_1l_2}     C_{l_2}^{EE}C_{l_3}^{E\alpha} \nn\\
&& + \mathcal{E}_{l_{\mathrm{sum}}} 8 h^{-1}_{l_1l_2l_3}  \sum_{L_2L_3L_1} \left\{ {\begin{array}{ccc}
		l_1 & l_2 & l_3  \\
		L_2 & L_3 & L_1  \\
\end{array} } \right\} (-1)^{L_1+L_2+L_3} \Bigg[I_{L_3L_1l_1}^{0-22}  I_{L_2L_1l_2}^{0-22}I_{L_2L_3l_3}^{0-22}    \mathcal{E}_{L_3L_1l_1}   \mathcal{E}_{L_2L_1l_2}   \mathcal{E}_{L_2L_3l_3}    \nn\\
&& +  \frac{1}{6}  \Big(I_{L_3L_1l_1}^{0-22} I_{L_1L_2l_2}^{0-22}I_{L_2L_3l_3}^{0-22}   \mathcal{E}_{L_3L_1l_1}   \mathcal{E}_{L_1L_2l_2}   \mathcal{E}_{L_2L_3l_3} \nn\\
&& + I_{L_1L_3l_1}^{0-22} I_{L_3L_2l_3}^{0-22}I_{L_2L_1l_2}^{0-22}   \mathcal{E}_{L_1L_3l_1}   \mathcal{E}_{L_3L_2l_3}   \mathcal{E}_{L_2L_1l_2}  \Big)C_{L_1}^{E\alpha}  C_{L_2}^{{E\alpha}}C_{L_3}^{{E\alpha}}  \Bigg]  + (5~ \mathrm{perms}),
\label{approximate_ppp} 
\eea

\acknowledgments

We thanks Gongbo Zhao for useful discussion on avoiding numerical instability in calculating the rotated bispectra, and thanks Chang Feng for review of the manuscript and suggestions on figure plotting. We also acknowledge
the use of CAMB package in calculating the power spectra and bispectra of CMB and polarization rotation. H. Z., S. L. , H. L. and X. Z. are supported in part by NSFC (Nos. 11653001, 11653003, 11653004), the Ministry of Science and Technology of China (2016YFE0104700),  and the CAS pilot B project (XDB23020000).  M. L. is supported by NSFC under Grants No. 11653002 and No. 11947301.


\end{document}